\documentclass[journal]{IEEEtran}
\usepackage{graphicx}
\usepackage{amsmath,amsfonts,amsthm,bm, bbm, amssymb}
\usepackage{psfrag}
\usepackage{color}
\usepackage{cite}
\usepackage{epsfig}
\usepackage{epstopdf}
\usepackage{graphicx}
\usepackage{lipsum}  
\usepackage{hyperref}
\usepackage{hypcap}
\usepackage[table]{xcolor}
\usepackage{wrapfig}
\usepackage[normalem]{ulem} 
\usepackage{multirow}
\usepackage[caption=false,font=normalsize,labelfont=sf,textfont=sf]{subfig}
\usepackage{algorithmic}
\usepackage{algorithm}
\usepackage{cancel}

\newcommand\norm[1]{\left\lVert#1\right\rVert}

\begin{document}
\title{Equitable 6G Access Service via Cloud-Enabled HAPS for Optimizing Hybrid  Air-Ground Networks 
}
\author{Rawan~Alghamdi,~\IEEEmembership{Graduate Student Member,~IEEE,}~
        Hayssam~Dahrouj,~\IEEEmembership{Senior Member,~IEEE,}~ 
        Tareq~Al-Naffouri,~\IEEEmembership{Senior Member,~IEEE,}~ 
        and~Mohamed-Slim~Alouini~\IEEEmembership{Fellow,~IEEE}~ 

}



\maketitle

\begin{abstract}
The evolvement of wireless communication services concurs with significant growth in data traffic, thereby inflicting stringent requirements on terrestrial networks. This work invigorates a novel connectivity solution that integrates aerial and terrestrial communications with a cloud-enabled high-altitude platform station (HAPS) to promote an equitable connectivity landscape.  Consider a cloud-enabled HAPS connected to both terrestrial base-stations and hot-air balloons via a data-sharing fronthauling strategy. The paper then assumes that both the terrestrial base-stations and the hot-air balloons are grouped into disjoint clusters to serve the aerial and terrestrial users in a coordinated fashion. The work then focuses on finding the user-to-transmitter scheduling and the associated beamforming policies in the downlink direction of cloud-enabled HAPS systems by maximizing two different objectives, namely, the sum-rate and sum-of-log of the long-term average rate, both subject to limited transmit power and  finite fronthaul capacity. The paper proposes solving the two non-convex discrete and continuous optimization problems using numerical iterative optimization algorithms. The proposed algorithms rely on well-chosen convexification and approximation steps,  namely, fractional programming and sparse beamforming via re-weighted $\ell_0$-norm approximation. The numerical results outline the yielded gain illustrated through equitable access service in crowded and unserved areas, and showcase the numerical benefits stemming from the proposed cloud-enabled HAPS coordination of hot-air balloons and terrestrial base-stations for democratizing connectivity and empowering the digital inclusion framework.
\end{abstract} 


\section{Introduction}
\IEEEPARstart{T}{he} ever-increasing demand for seamless connectivity, high capacity, and wider wireless coverage has driven the recent drastic evolution of wireless communication networks. 
Despite the massive advancements achieved through the previous generations of wireless systems, approximately four billion people remain unconnected \cite{yaacoub_6G}, most of whom reside in rural areas, underdeveloped areas, and less-developed countries and islands. Achieving adequate wireless connectivity and broadband coverage has been one of the fundamental challenges to wireless communication providers. Such a challenge is accentuated in the present era characterized by the prolific growth of connected devices and data deluge. Hence, while the past five commercial generations of wireless communication systems (i.e., 1G-5G) have focused on improving capacity-related metrics, the communication research community envisions the next generation to focus on democratizing connectivity and information and communications technologies (ICTs) as a universal public utility \cite{yaacoub_6G}. To this end, this work offers one step forward toward providing a ubiquitous connectivity solution anywhere, anytime, by considering a connectivity-from-the-sky platform empowered by cloud computing capabilities.

The classical terrestrial network (TN)-based connectivity solution relies on over-engineering and network densification to address the increasing demands; however, such solutions are neither sustainable nor sufficient for the demands of future networks, especially since network densification results in a drain of spatial resources, performance degradation, and significant energy consumption \cite{liu_densification}.  Providing a ubiquitous, robust, and balanced wireless service, therefore, requires a paradigm shift in designing future networks \cite{dang_6G}. A recent trend in the wireless communication community is the integration of TN with non-terrestrial networks (NTN) to form a vertical heterogeneous network (VHetNet) \cite{alzenad_vhetnet}. Along this direction, this article proposes a cloud-enabled high-altitude platform station (HAPS) system architecture that offers joint communication and computation for both aerial and terrestrial users.
 The proposed cloud-enabled HAPS provides a high-speed data-sharing fronthaul connection to aerial (hereafter denoted by hot-air balloons) and terrestrial BSs. Furthermore, the proposed system model groups the hot-air balloons and terrestrial BSs into mutually exclusive clusters, where the BSs within a cluster  coordinate the resource management and multipoint transmission techniques to provide access to terrestrial and flying users. To this end, the paper focuses on two distinct system performance optimization problems by optimizing the user-to-transmitter scheduling and the corresponding beamforming vectors, which control the level of interference experienced by both flying and ground users in VHetNets.

\subsection{Related Work}
The optimization problems tackled in the current paper are related to the beamforming and user-scheduling design in VHetNets, particularly in cloud-enabled HAPS systems.  While the beamforming and scheduling problems are studied extensively in the recent wireless literature (both jointly and separately), their direct applications to the multi-layered cloud-enabled VHetNets specifics remain relatively limited.  This section next overviews the general resource management applications to VHetNets and cloud-enabled systems, and specifies the major differences between the classical resource optimization problem and the problems considered in the current paper.

The realization of VHetNets allows the communication architecture to expand both in breadth (through extended coverage) and in-depth (through serving users at different heights), which provides the network with extra layers of connectivity, flexibility,  agility, and resilience. NTNs often consist of airborne and spaceborne (satellite) communication. The deployment of standalone space-ground communication remains relatively expensive, and the path loss of satellite-to-earth links is severe. The associated space-earth inter-communication also often incurs huge delays. In contrast, the deployment of airborne stations has lower cost, reduced delay, and more favorable line-of-sight compared to TNs \cite{cao_aircomm}. The aerial layer of NTNs consists of  both
 HAPS and low-altitude platforms (LAPs) (e.g., unmanned air vehicles (UAVs)). Previous works in the literature study UAV-assisted wireless communication networks, where the UAV act as a flying base-station (BS) \cite{zhang_uav}. Although such networks address some of the demands of the sixth generation of wireless networks (6G) and beyond, UAVs are energy-limited platforms and thus have short endurance. Moreover, UAVs fly in low altitudes and have a relatively small wireless footprint  \cite{cao_aircomm}.

On the other hand, HAPS, often deployed as quasi-stationary modules in the stratospheric layer, can provide high capacity and improved coverage while offering a longer flying duration \cite{kurt_vision}. HAPS,  therefore, emerges nowadays as a vital network component of VHetNets,  integrating the terrestrial and aerial layers seamlessly and improving the network's flexibility, resilience, and robustness. HAPS has particularly emerged recently 
as a key enabler for achieving the ubiquitous connectivity goals of 6G networks \cite{kurt_vision}. Given its strategic hovering altitude,  HAPS is considered the medium infrastructure that seamlessly connects space-borne, airborne, and terrestrial communication systems. More specifically, HAPS-assisted networks support mobile and fixed users, enabling the networks to satisfy different service demands. HAPS-assisted networks are economical solutions that provide large capacity to hyper-digitized areas while ensuring connectivity and coverage to under-served and rural areas \cite{mershad_cloudhaps}. HAPS-assisted networks indeed have the potential to improve the accessibility and quality of communication services by complementing and consolidating TNs \cite{kurt_vision, alam_HAPS_SMBS}.

   In addition to the capability of HAPS to elevate the capacity and coverage of wireless communication systems, HAPS systems can enable computing, caching, sensing, and control. The recent literature features such an interplay of communication and computation and further explores the opportunities stemming from integrating computing services in the HAPS. For example, the work in  \cite{alzenad_fso} demonstrates the powerful performance of vertical front/backhaul links using HAPS to provide high-speed free-space optics (FSO) backhaul links to small cells on the ground. Furthermore, the work in  \cite{mershad_cloudhaps} is the first to propose a cloud-enabled HAPS system, where the authors explore the opportunities and potential applications of enabling cloud computing, edge computing, data offloading, and front/backhauling from the sky. Moreover,  \cite{mershad_cloudhaps} describes the prospects of a cloud-enabled HAPS system in integrating traditional cloud services, among other new possibilities for cloud services, such as management services for satellites, sensors, and transportation. In fact, cloud-enabled HAPS systems allow the network to have smooth scalability, where TNs offload traffic data to the more computationally powerful aerial node (i.e., HAPS) for data processing while allowing a simple design on the ground.

 The deployment of the proposed cloud-enabled HAPS system, as any traditional heterogeneous cloud-radio access network (HC-RAN),  is subject to specific physical constraints,  i.e., depending on the availability of communication resources and the limited level of coordination between the BSs \cite{emil_multicell}.  Generally speaking,  resource allocation in interference networks is challenging due to the non-convexity of the majority of interference management problems; hence, the literature focuses on devising practical numerical solutions for optimizing  HC-RANs.  For instance, various techniques for local optimality have been proposed over the years with different considerations and limitations. For example, the authors in \cite{gesbert_multicell} present a comprehensive overview of interference management techniques for cooperative multiple-input multiple-output (MIMO) systems, where several BSs within the same cell cooperate to serve users. The authors in \cite{gesbert_multicell}  highlight the constraints on such a network MIMO approach in terms of limited backhaul capacity, per-BS power consumption, and the need for decentralized processing; some of such challenges are further addressed in \cite{chowdhery_multicell, yu_multicell}.

 The work in \cite{chowdhery_multicell} examines the problem of maximizing a network-wide utility in a downlink multi-cell wireless orthogonal frequency division multiple-access (OFDMA) system subject to a finite backhaul capacity (cooperation between BSs). The authors in \cite{chowdhery_multicell} propose zero-forcing beamforming per frequency carrier and formulate the objective function to address the backhaul capacity constraint as a penalty term.  In order to integrate the multi-fold resource allocation, the authors propose an iterative solution that alternates between solving the beamforming vectors and user scheduling policy. On the other hand, the work in \cite{yu_multicell} considers maximizing the network utility subject to power constraints on a per-BS per-frequency tone basis by jointly and iteratively determining user scheduling, beamforming, and power spectrum adaptation techniques in heterogeneous networks. 
 In addition, the works in \cite{chowdhery_multicell} and \cite{yu_multicell} focus on maximizing a network-wide utility to ensure a load-balancing framework among users. The optimization problems featured in \cite{chowdhery_multicell} and \cite{yu_multicell} consider the fair distribution of loads among users by taking the utility function as the logarithm of the average rate of users over a long period. Such a choice of optimization objective is shown to be optimal for distributing the rate gains proportionally among users \cite{load_andrews}.

 Unlike the above works which assume that the association of users to the set of serving BSs is static, adopting a dynamic user association approach allows for more degrees of freedom and ensures connectivity and stable services for cell-edge users \cite{dai_cran}.   In dynamic user clustering, the resource allocation per BS can be challenging as dynamic clustering requires more signaling overhead to establish new associations continuously. For example, the work in \cite{dai_cran} considers a joint clustering and beamforming design for maximizing network utility in HC-RAN under finite backhaul capacity per-BS constraints. Instead of using the backhaul constraint as a penalizing term as in \cite{chowdhery_multicell}, the authors in \cite{dai_cran} propose using the $\ell_1$-norm re-weighting technique for the constraint, coupled with the generalized weighted minimum mean square error (WMMSE) beamforming technique  \cite{shi_wmmse, wmmse}. Similarly, the work in \cite{ahmad_beamforming} uses the proposed $\ell_1$-norm re-weighting technique for finite backhaul constraints and solves a sum-rate maximization problem over the beamforming and clustering strategies. By tackling the backhaul constraint using the $\ell_1$-norm re-weighting technique, the authors in \cite{ahmad_beamforming} use fractional programming techniques \cite{FP1, FP2} to solve a sum-rate maximization problem.

While the resource management challenges in cloud-enabled HAPS systems are similar to those in HC-RAN systems, classical resource management techniques must be carefully adapted to VHetNets to account for connectivity constraints stemming from the three-dimensional space and the physical limitations of the aerial platforms. The recent literature features a limited number of works tackling the resource management problems in VHetNets. For example, the works \cite{alsharoa, Liu_SatHAPS, ding_hapsedge,  Pan_SAGIN} consider a resource management framework for satellite-HAPS-ground VHetNet. Specifically,  the optimization frameworks in \cite{ding_hapsedge} and \cite{Pan_SAGIN} study energy management in VHetNets. The work in \cite{ding_hapsedge} presents a resource management optimization problem to jointly minimize the sum of energy over the user communication and computation assignment, beamforming vectors, and computation resource allocation assignment. Similarly, the work in \cite{Pan_SAGIN} presents an energy efficiency maximization problem in satellite-HAPS-ground VHetNets, where the authors analyze the system performance of satellite-HAPS-ground cooperative relay and maximize its end-to-end energy efficiency over the transmit power and time allocation.  On the other hand, the resource management schemes in \cite{alsharoa} and \cite{Liu_SatHAPS } address the network throughput maximization via optimizing the user access association to the satellite, HAPS, or terrestrial BSs, transmit power, and backhaul capacity constraints in \cite{Liu_SatHAPS}, and optimizing the backhaul and access association, transmit power and HAPS' trajectory in \cite{alsharoa}.

Along the same line, this paper aims to maximize the network throughput and user fairness by jointly optimizing the beamforming vectors and user association in a cloud-enabled HAPS system. Unlike the above works, our paper considers the cooperation of terrestrial BSs and hot-air balloons in disjoint clusters to serve both aerial and terrestrial users, the connectivity constraint of terrestrial BSs to serve aerial users, and finally, the physical resource limitation characterized by the per-BS fronthaul capacity and transmit power.

\subsection{Contribution}

Given the various advantages of HAPS systems in providing connectivity from the sky, this article investigates the role of HAPS as a cloud computing and communication platform (i.e., cloud-enabled HAPS). The proposed architecture provides a cost-effective connectivity solution to users in rural and isolated regions while ensuring a stable connection to users in urban areas \cite{mershad_cloudhaps}. 
Moreover, the HAPS cloud in our work is assumed to be connected to disjoint clusters of both hot-air balloons and terrestrial BSs via high-speed yet finite FSO fronthaul links, where the coordination of the mutual transmission strategies occurs across the BSs within one cluster. Our work further assumes that aerial users (e.g., LAPS, drones) are exclusively served by the hot-air balloon clusters, which is a practical design consideration since serving aerial users using terrestrial BSs requires full-dimensional antenna arrays \cite{zeng_uavUE, mozaffari_uavUE}. Terrestrial users, on the other hand, can be served by hot-air balloon clusters or terrestrial BS clusters. 
Assuming a spatial multiplex system in the downlink direction using beamforming, this work optimizes the user-to-transmitter association strategy, jointly with the beamforming vectors, subject to per-BS fronthaul capacity and transmit power constraints. In particular, this article studies two different optimization problems inspired by the need to create a digital equity paradigm.  We next highlight the contributions of each considered problem in its own right.

\begin{enumerate}
\item The first optimization problem aims at maximizing the network sum-rate subject to physical design constraints characterized by the limited transmit power and fronthaul capacity by means of jointly determining the association of users to a cluster of BSs and their corresponding beamforming vectors. The major contributions considered in the problem are as follows:

\begin{itemize}
	\item We formulate a sum-rate maximization problem over the user-to-transmitting clustering and the corresponding beamforming vectors, subject to connectivity constraints, limited transmission power, and finite fronthaul capacity. 
	\item We tackle the non-convex mixed discrete-continuous optimization intricacies using convex approximation and reformulation techniques. 
	\item We use sparse beamforming to relax the non-convex identity function in the fronthaul capacity constraints using re-weighted $\ell_0$-norm approximation. 
	\item We adopt fractional programming techniques to convexify the objective function via employing the corresponding quadratic transform to recast sum-log-ratio into a convex sum-ratio. 

	\item We propose a numerical optimization algorithm that iteratively designs a user clustering strategy that judiciously associates a set of mutual users to a cluster, limiting excessive interference, then determines the corresponding beamforming vectors for the active user set. 
	
	\item We then assess the performance of the proposed solution by demonstrating the significant gain obtained by augmenting terrestrial wireless communication services with airborne communication supported by a fronthaul connection from the cloud-enabled HAPS to both the hot-air-balloons and the terrestrial BSs.

\end{itemize}

\item While the sum-rate maximization problem considered in this work shows significant gains in yielded throughput, our numerical results show that such a greedy optimization framework tends to favor users with good channel gains, leaving others unserved.  Such a performance motivates looking into a transmission scheme that exploits multi-user diversity gains by allocating different resource blocks to different users, aiming to combat rate disparity between users. Hence, the second optimization framework considered in this paper focuses on devising a connectivity framework that maximizes the summation-of-logarithm (sum-log) of the average rate over a long period in an OFDMA-enhanced cloud-enabled HAPS system. 

The main idea of maximizing the logarithm of a utility is that it naturally achieves a fair, load-balanced scheme among users. The major contributions considered in the problem are then as follows:

\begin{itemize}
	\item We consider a multi-user (multiple-access) diversity scheme in a  cloud-enabled HAPS system and schedule users over the appropriate resource blocks.
	\item We demonstrate that the problem of creating a proportionally fair service becomes intertwined with both the assignment of users to the available resource blocks within a cluster and the design of the respective beamformers.
	\item We jointly optimize the scheduling of users to a set of frequency tones (i.e., resource blocks) and beamforming vectors to maximize the sum-log of the average rate over a long period subject to physical design constraints characterized by the limited transmit power and fronthaul capacity in an OFDMA scheme to obtain a proportionally fair access service.
	\item  We propose a numerical iterative optimization algorithm, which tackles the optimization problem over one parameter at a time. 	
	\item The proposed solution shows that employing the proportional fairness criterion transforms the sum-log of the average rate objective function into a weighted sum-rate.
	\item The proposed optimization framework iteratively solves a weighted sum-rate problem that considers practical system constraints, namely per-BS fronthaul capacity and transmit power constraints.

	\item We adopt sparse beamforming and fractional programming techniques to solve the intricate non-convex mixed integer optimization problem. Notably, we utilize the quadratic transform to recast the sum-log of the average rate into the sum-log of ratios by finding a numerical algorithm solution to solve the problem of interest.
	
	\item Finally, we provide numerical results that illustrate the gain in user fairness obtained from cloud-enabled HAPS systems, which portray the role of the proposed model in providing an equitable access service to span both hyper-digitized and unconnected areas without sacrificing throughput.  
	\end{itemize}

\end{enumerate}

\subsection{Organization}
 In what follows, the article discusses the proposed cloud-enabled HAPS system model in Section \ref{sec:system_model}. Then, the article features two distinct optimization problems in Section \ref{sec:sumrate} and Section \ref{sec:sumlograte}.  Specifically, Section \ref{sec:sumrate} considers the sum-rate maximization problem and proposes a joint user-association and beamforming design algorithm. Section \ref{sec:sumlograte} emphasizes the importance of creating a fair service cloud-enabled HAPS and considers a sum-log rate problem in a  multiple access scheme that can achieve a fair and balanced service under the joint optimization of user scheduling and beamforming design. Section \ref{sec:results} provides numerical results for the proposed algorithms. Section \ref{sec:conc} finally concludes the paper.

\section{System Model}\label{sec:system_model}
\begin{figure}
    \centering
    \includegraphics[width=\linewidth]{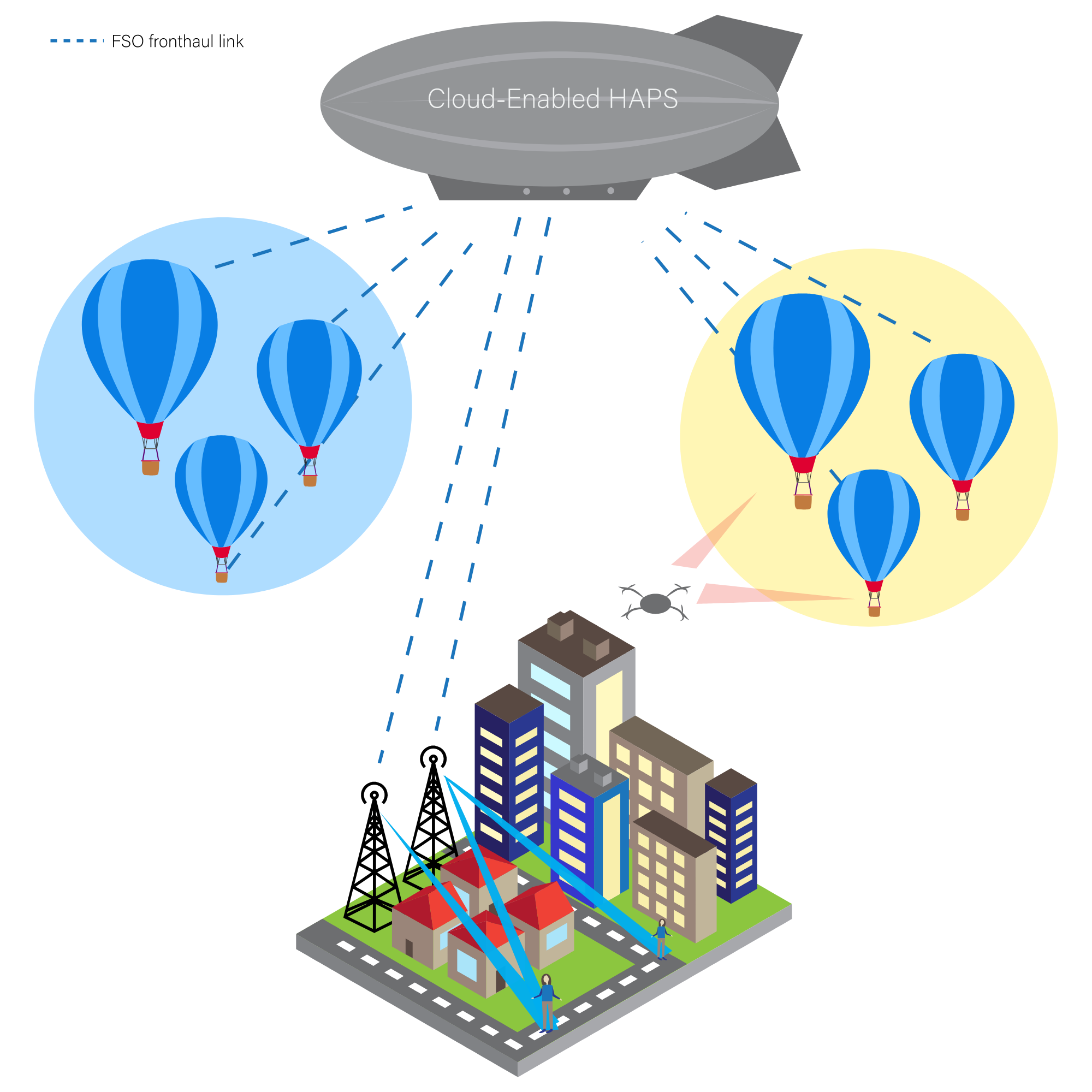} 
    \caption{Cloud-enabled HAPS system model highlighting a network-MIMO approach of serving users.     \label{fig: systemch3}}
\end{figure}

This article considers a multilayered cloud-enabled HAPS network, similar to Fig.~\ref{fig: systemch3}, consisting of $B_A$ hot-air balloons (HBSs) and $B_T$ terrestrial BSs  (TBSs) connected to one cloud-enabled HAPS via high-speed FSO fronthaul links. All TBSs and HBSs are assumed to have limited transmit power and fronthaul capacity. Connecting all  TBSs and HBSs to the HAPS cloud processor allows them to cooperate to serve all users. Although this approach reduces the interference drastically, it requires high fronthaul capacity, which can be physically infeasible \cite{dai_cran}. Therefore, we consider grouping the $B = B_T + B_A$ BSs, each equipped with $M$ antennas, into $Q$ mutually exclusive clusters; each cluster consists of $L$ BSs, where the BSs within each cluster can coordinate to serve their associated users in a network MIMO fashion to control the interference in the network. We classify the clusters as $Q_A$ hot-air balloons clusters and $Q_T$ terrestrial clusters.  The grouping of BSs into mutually exclusive clusters is assumed to be fixed, and we denote the set of BSs in cluster $q$ by $\mathcal{B}_q$, where $ \cup_{q=1}^Q\mathcal{B}_q = \mathcal{B} $. Let  $\mathcal{Q} = \mathcal{Q}_A \cup \mathcal{Q}_T$ denote the set of all  TBSs and HBSs clusters, where hot-air balloons and terrestrial sets are denoted by $\mathcal{Q}_A$ and $ \mathcal{Q}_T$, respectively.  Moreover, the paper assumes that there are $U = U_T + U_A$ single-antenna users, where $U_T$ is the number of terrestrial users and $U_A$ is the number of aerial/UAV users. The users are assumed to be separated via spatial multiplexing, i.e., through beamforming. Similarly, we denote the set of aerial and terrestrial users by $\mathcal{U}_A$ and $\mathcal{U}_T$, respectively, and their union by $\mathcal{U}$.

Furthermore, this work assumes that the cloud-enabled HAPS has perfect channel state information of the entire network. Further,  the cloud processor performs system-level optimization and aggregates the beamforming vectors and other related information to the  TBSs and HBSs using FSO fronthaul links. Finally, the work assumes that the communication between the users and both TBSs and HBSs(i.e., access links) operate at the radio frequency (RF) range. Hence, there is no interference between the access and fronthaul links; the system performance is rather limited by both the intra- and inter-cluster interference.
 Let $\mathbf{w}_{u,b,q} \in \mathbb{C}^{M \times 1}$ be the beamforming vector of BS $b$ in cluster $q$ serving the associated user $u$. Furthermore, let 
 $\mathbf{w}_{u,q} \in \mathbb{C}^{LM\times 1}$ be the aggregate transmission beamforming vector at cluster $q$ for user $u$.

The signal received by user $u$ served by cluster $q$ is then expressed as
\begin{equation}
y_{u,q} =  \mathbf{h}_{u,q}^{\dagger}\mathbf{w}_{u,q} z_{u,q} s_u + \sum_{\substack{(i,j) \neq (u,q), \\ i \in \mathcal{U}, j \in \mathcal{Q}}} \mathbf{h}_{u,j}^{\dagger} \mathbf{w}_{i,j}z_{i,j} s_i +\nu_u, 
\label{eq:y}
\end{equation}
\noindent where $ \mathbf{h}_{u,q} \in \mathbb{C}^{ML \times 1}$ denotes the aggregate channel vector from all of the BSs in cluster $q$ to user $u$, the operator $(\cdot)^{\dagger}$ denotes the Hermitian transpose, $z_{u,q}$ is a binary variable denoting the association of user $u$ to cluster $q$, $\nu_i \sim \mathcal{CN}(0,\sigma^2)$ is the noise at user $i$. The notation $s_i \sim \mathcal{CN}(0,1)$ denotes the signal intended for user $i\in \mathcal{U}$, where we assume that data streams between distinct users are independent and identically distributed (i.i.d). 

 Let user $u$ be associated with cluster $q$, then using \eqref{eq:y} the signal-to-interference-plus-noise ratio (SINR) can be defined as
 \begin{equation}
 \mathrm{SINR}_{u,q} =\frac{z_{u,q}\big|{\mathbf{h}}_{u,q}^\dagger  {\mathbf{w}_{u,q}} \big|^2}{\sum_{\substack{(i,j) \in \mathcal{U}\times \mathcal{Q},\\ (i,j)\neq (u,q)}}{z_{i, j}\big|\,{\mathbf{h}}_{u,j}^\dagger {\mathbf{w}_{i,j}} }\big|^2 + \sigma^2},
 \label{eq:sinr}
 \end{equation}

 It is worth noting that the summation term in the denominator of \eqref{eq:sinr} considers both inter-cluster (i.e., $j\neq q$) and intra-cluster (i.e., $j = q$) interference. The inter-cluster interference describes the power of the interfering signals from other clusters when serving other users. On the other hand, intra-cluster interference describes the power of interfering signals due to serving other users within the same cluster. The achievable rate (throughput) of user $u$ when associated with clustering $q$ is then expressed as
\begin{equation}
R_{u,q} = \log_2(1+\mathrm{SINR}_{u,q}). \label{eq:rate}
\end{equation}

Given \eqref{eq:rate}, we can see that the parameters affecting the obtained network-wide throughput are the clustering choice and the design of the beamforming vectors. 
Equally important, the network throughput is limited by the available network resources, namely power and capacity. The following two sections formulate two different optimization problems over the user-to-transmitter scheduling and association and the beamforming vectors subject to the network's finite resources.

\section{Throughput Maximization via Beamforming and User Association}\label{sec:sumrate}

This section formulates an optimization problem that exploits the interrelation of the system performance and the available resources per transmitter (i.e., limited transmit power and fronthaul capacity). Namely, this section tackles an optimization problem that maximizes a network-wide sum-rate, subject to resource availability constraints, by jointly considering user association and the respective beamforming vectors. The presented problem features a non-convex optimization problem over mixed discrete (user-association) and continuous (beamforming) optimization variables. This section tackles the intricacies of the optimization problems using well-chosen mathematical reformulation techniques, such as sparse beamforming using both the $\ell_1$-norm re-weighting technique, and fractional programming (FP) transformations, first proposed in \cite{FP1, FP2}. 

\subsection{Problem Statement}\label{sec:sumrate_prob}
The sum-rate maximization problem considered in this paper is a function of both the beamforming vectors and the user clustering choice. 
  Motivated by the performance improvement of jointly optimizing user association and their corresponding beamforming vectors compared to tackling them separately, this section studies the joint optimization of user assignment and beamforming, and demonstrates an iterative approach to solve the optimization problem.

 Given the system model in Section \ref{sec:system_model}, we formulate a sum-rate maximization problem constrained by per-BS transmit power, fronthaul capacity, and practical connectivity constraints. In particular, the connectivity constraints dictate that every user can only be associated with one and only one transmitting cluster. Furthermore,  we assume  TBSs cannot serve aerial users, which is a practical design consideration since serving aerial users using  TBSs requires full-dimensional antenna arrays  \cite{zeng_uavUE, mozaffari_uavUE}.  The first optimization framework can then be formulated as follows: 

 \begin{subequations}
    \begin{align}
    \max_{\mathbf{w}, \mathbf{z}, \mathbf{r}} \quad & \sum_{ q \in  \mathcal{Q}}\sum_{u \in \mathcal{U}} R_{u, q}    \label{opt_ch3:obj}  \\
    \textrm{s.t.}&~ (\ref{eq:rate}) \\
    			 &\sum_{q \in  \mathcal{Q}} z_{u, q} \leq 1, ~\forall u \in \mathcal{U} \label{opt_ch3:const2}\\
    			 & \sum_{q \in \mathcal{Q}_T}\sum_{u \in \mathcal{U}_A} z_{u, q} = 0, \label{opt_ch3:const3}\\
    			 &\sum_{u \in \mathcal{U}} ||\mathbf{w}_{u,b,q}||_2^2 z_{u, q} \leq  P_{b,q},  ~ \forall b \in \mathcal{B}_q, \forall q \in \mathcal{Q}\label{opt_ch3:const4}\\
    			 &\sum_{u \in \mathcal{U}} R_{u, q} \mathbbm{1} \{||\mathbf{w}_{u,b,q}||_2^2\} \leq C_{b, q},\forall b \in \mathcal{B}_q, \forall q \in \mathcal{Q}.\label{opt_ch3:const5}
    \end{align} \label{opt_ch3:Opt}
 \end{subequations}

The optimization problem in \eqref{opt_ch3:Opt} is over the user-to-transmitter association vector $\mathbf{z} = \{z_{u,q} | (u, q) \in \mathcal{U} \times \mathcal{Q} \}$, the set of beamforming vectors $\mathbf{w} = \{\mathbf{w}_{u,q} | (u, q) \in \mathcal{U} \times \mathcal{Q} \}$, and the rate of all links $\mathbf{r} = \{R_{u, q} | (u, q) \in \mathcal{U} \times \mathcal{Q}\}$. The objective function, (\ref{opt_ch3:obj}),  maximizes users' sum-rate. Constraint (\ref{opt_ch3:const2}) ensures that every user is served by one cluster at most, while (\ref{opt_ch3:const3}) states that the aerial users (UAV-user equipment) 
are exclusively served by HBS clusters.
For the  $b$-th BS in the $q$-th cluster, constraint (\ref{opt_ch3:const4}) represents its maximum transmission power, $P_{b,q}$, and (\ref{opt_ch3:const5}) represents its maximum fronthaul capacity, $C_{b, q}$. 

 Apart from the non-convexity of the objective function, the fact that the optimization variables in \eqref{opt_ch3:Opt} are a mix of continuous (beamforming) and discrete (association) variables makes the problem a complex one. We approach the intricacies of the problem using FP and sparse beamforming approaches, as this section illustrates. More specifically, this section highlights the challenges of solving the problem jointly for the association binary variables and the beamforming vectors, studies the reformulation of the problem to simplify the constraints, and proposes possible methods to untangle the intricacies of the formulated problem. 

\subsection{Problem Transformation}\label{sec:sumrate_reform}

The optimization problem \eqref{opt_ch3:Opt} is a complicated problem and cannot be remodeled as a convex problem because of the mixed nature of the optimization variables. This section tackles the problem by first simplifying the equality constraint in \eqref{opt_ch3:const3}. Since the interference experienced by one user is a function of the clustering decision of other users, we claim that the constraint in \eqref{opt_ch3:const3} is implied implicitly in the interference term of \eqref{eq:rate}. In other words,  the rate and SINR constraints themselves take into account that the  TBSs  are not serving the aerial users. Generally speaking, terrestrial users may receive interfering signals from hot-air balloons clusters that serve other terrestrial users or from terrestrial clusters serving other users. In contrast, aerial users may experience inter-cluster interference from hot-air balloon clusters serving users, but the signals from terrestrial clusters would not leak at 
the flying users' receivers.

 Another intricacy in the formulation of \eqref{opt_ch3:Opt} is the discrete indicator function in constraint \eqref{opt_ch3:const5}. This indicator function characterizes whether user $u$ is served by the $b$-th BS in cluster $q$,  i.e., depending on whether the beamforming vector of the link between $u$ and $q$ is nonzero. Hence, the fronthaul capacity of BS $b$ is the summation of the rates of the users served (or cooperatively served) by BS $b$. To make this constraint tractable, we propose relaxing the indicator function in constraint \eqref{opt_ch3:const5} using a sparse beamforming approach, as in \cite{dai_cran} and \cite{ahmad_beamforming}. 
First, we observe  that the argument of $ \mathbbm{1} \{\cdot\}$ in \eqref{opt_ch3:const5} is a positive real number (i.e., $||\mathbf{w}_{u,b,q}||_2^2 \in \mathbb{R}^+$), and so the  indicator function reduces to the $\ell_0$-norm. Then, we adopt the weighted $\ell_0$-norm approximation to convexify the $\ell_0$-norm \cite{dai_cran, candes_l1}. 

 Therefore, we can rewrite the indicator function in \eqref{opt_ch3:const5} as 
 \begin{equation}
    \mathbbm{1} \{||\mathbf{w}_{u,b,q}||_2^2\} \triangleq \norm{ ||\mathbf{w}_{u,b,q}||_2^2}_0 \approx \beta_{u,b,q} ||\mathbf{w}_{u,b,q}||_2^2, \label{eq:ell0}
 \end{equation}
 \noindent where $\beta_{u,b,q}$ is  a constant weight associated with $||\mathbf{w}_{u,b,q}||_2^2$ to approximate the $\ell_0$-norm. The weight   $\beta_{u,b,q}$ is updated iteratively with a regularization factor $\epsilon > 0$, given the following rule
 \begin{equation}
    \beta_{u,b,q} = \frac{1}{\epsilon + ||\mathbf{w}_{u,b,q}||_2^2}, \label{eq:beta}
 \end{equation}
 \noindent which corresponds to the inverse of the power allocated by BS $b$ to serve user $u$. The weight updating rule is designed to discourage the set of  BSs with low transmit power from serving user $u$, such that only BSs with a considerable contribution to the rate of user $u$ are encouraged to cooperate.

 Given the two simplifications above, we reformulate problem \eqref{opt_ch3:Opt} as 
 \begin{subequations}
    \begin{align}
    \max_{\mathbf{w}, \mathbf{z}, \mathbf{r}} \quad & \sum_{ (u,q) } R_{u, q}    \label{opt_ch3:reobj}  \\
    \textrm{s.t.}&~ (\ref{eq:rate}), \eqref{opt_ch3:const2}, \eqref{opt_ch3:const4}\\
    			 &\sum_{u \in \mathcal{U}} R_{u, q} \beta_{u,b,q} ||\mathbf{w}_{u,b,q}||_2^2 \leq C_{b, q}.~  \label{opt_ch3:reconst5}
    \end{align} \label{opt_ch3:reopt}
 \end{subequations}

 The reformulation of the original problem presented in \eqref{opt_ch3:reopt} remains difficult to solve, mainly because the rate term $R_{u, q}$, which appears in \eqref{opt_ch3:reobj} and \eqref{opt_ch3:reconst5}, is not convex. In what follows, we propose solving the problem iteratively with the help of FP theory \cite{FP1, FP2} to tackle the non-convexity and mixed variables.

\subsection{Proposed Solution}\label{sec:sumrate_sol}
 In this section, we apply FP techniques to decouple the numerator and denominator of the SINR term in \eqref{opt_ch3:reopt}, which are both functions of the two optimization variables of interest. 
Although \eqref{opt_ch3:reobj} features a summation of the logarithm of ratios instead of a summation of ratios, the recent advances in FP theory, namely the quadratic transform in \cite{FP1, FP2}, allow us to apply FP to a function of a ratio as long as the function is nondecreasing and concave. Hence,  this section applies the quadratic transform approach to recast 
\eqref{opt_ch3:reopt}  into a series of convex optimization problems and solve them iteratively.

 As Section \ref{sec:sumrate_reform} discusses, the non-convex constraint \eqref{opt_ch3:reconst5} makes the problem difficult to solve. Hence, we propose solving \eqref{opt_ch3:reopt} iteratively by fixing the rate expression with $\tilde{R}_{u, q}$ in \eqref{opt_ch3:reconst5}, and simultaneously updating $\tilde{R}$ and $\beta$. By doing so, \eqref{opt_ch3:reconst5} appears to be a weighted power constraint, which is quadratic (hence convex) in $\mathbf{w}$ \cite{dai_cran}. Given these points, we can express the sum-rate maximization problem as
 
 \begin{subequations}
    \begin{align}
    \max_{\mathbf{w}, \mathbf{z}, \mathbf{r}} \quad & \sum_{ (u,q) } R_{u, q} \label{opt_ch3:objectivR}  \\
    \textrm{s.t.}&~ (\ref{eq:rate}), \eqref{opt_ch3:const2}, \eqref{opt_ch3:const4}\\
    			 &\sum_{u \in \mathcal{U}} \tilde{R}_{u, q} \beta_{u,b,q} ||\mathbf{w}_{u,b,q}||_2^2 \leq C_{b, q}.
    \end{align}
 \end{subequations} 
 
 The remaining numerical hurdles at this stage are related to the mixed optimization variables and the non-convex objective problem. Next, we tackle the objective function using FP, in particular, the quadratic transform theorem \cite{FP1}. Although \eqref{opt_ch3:objectivR} has a sum-of-log-of-ratio form, we argue that because the main component of the objective function is a fraction (i.e., the SINR term in the $R_{u, q}$), we can apply FP. Particularly, we apply multidimensional and complex FP quadratic transform \cite[Theorem 2, Theorem 3]{FP1}. 
 
 The idea is---by exploiting Lagrange's dual reformulation of the objective function---we can remodel the objective function from a sum-of-log to a sum-of-ratio form, which enables the use of the quadratic transform.

 We introduce a new optimization variable $\gamma_{u, q}$. That is, we have

\begin{subequations}
\begin{align}
\max_{ \mathbf{w},  \mathbf{z},  \mathbf{\gamma}} \quad & \sum_{(u,q) \in \mathcal{U} \times \mathcal{Q}}    \log_2 \left( 1 + \gamma_{u, q} \right)    \\
\textrm{s.t.}&~ \gamma_{u, q} \leq \mathrm{SINR}_{u,q}, ~\forall (u,q) \in \mathcal{U} \times \mathcal{Q} \label{eq: gamma} \\
			 &\sum_{q \in  \mathcal{Q}} z_{u, q} \leq 1, ~\forall u \in \mathcal{U} \\
			 &\sum_{u \in \mathcal{U}} ||\mathbf{w}_{u,b,q}||_2^2 z_{u, q} \leq  P_{b,q}, \\
			 &\sum_{u \in \mathcal{U}} \tilde{R}_{u, q} \beta_{u,b,q} ||\mathbf{w}_{u,b,q}||_2^2 \leq C_{b, q}.
\end{align}  \label{opt_ch3:gamma_prob}
\end{subequations}

The approach is then to solve problem \eqref{opt_ch3:gamma_prob} in two stages (i.e., loops). We optimize over $ (\mathbf{w}, \mathbf{z})$ in the outer loop and over the set of auxiliary variables $\mathbf{\gamma}$ in the inner loop.
 
The latter problem is convex in $\mathbf{\gamma}$, and hence the strong duality theorem applies and has a trivial solution in which the inequality in \eqref{eq: gamma} is satisfied by equality. In other words, 
\begin{equation}
    \gamma_{u, q}^{\ast} =\mathrm{SINR}_{u,q}.  \label{eq: gamma_star}
\end{equation}

We then exploit the duality of the inner problem and find the partial Lagrangian function as

\begin{align}
\mathcal{L}(  \mathbf{\gamma},  \mathbf{\mu}) &= \sum_{(u,q) \in \mathcal{U} \times \mathcal{Q}}   \log_2 \left( 1 + \gamma_{u, q} \right) - \mu_{u, q} \left( \gamma_{u, q} - \mathrm{SINR}_{u,q} \right),  
\label{eq: lagrange}
\end{align} 
\noindent where $ \mathbf{\mu} = \{ \mathbf{\mu}_{u,q} | (u, q) \in \mathcal{U} \times \mathcal{Q} \}$ is the Lagrangian dual multiplier. 
Next, we set the partial derivative of (\ref{eq: lagrange}) for $ \gamma_{u, q}$ to zero, and solve for $ \gamma_{u, q}$ to yield:

\begin{subequations}
\begin{align}
 \mu_{u, q}^{\ast} = 
 \frac{1 }{1 + \gamma_{u, q}^{\ast}} \label{eq: lambda} 
\end{align} 
\end{subequations}\label{eq: first_order}

Here, $( \gamma_{u, q}^{\ast},  \mu_{u, q}^{\ast})$ are the saddle points of the dual problem of \eqref{eq: lagrange}, where $\gamma_{u, q}^{\ast}$ is the trivial solution of \eqref{opt_ch3:gamma_prob} found in \eqref{eq: gamma_star}. By substituting  $ \mu_{u, q}^{\ast}$ in (\ref{eq: lagrange}), we arrive at the following inner problem

\begin{equation}
\max_{ \mathbf{\gamma}} \quad \mathcal{L}(  \mathbf{\gamma},  \mathbf{\mu}^{\ast}).
\end{equation} \label{opt_ch3:dual}

Now, we combine the inner problem with the outer problem to get

\begin{subequations}
\begin{align}
\max_{ \mathbf{w},  \mathbf{z},  \mathbf{\gamma}} \quad & f_r(\mathbf{w}, \mathbf{z}, \mathbf{\gamma})   \\
\textrm{s.t.}&~ \sum_{q \in  \mathcal{Q}} z_{u, q} \leq 1,\\
			 &\sum_{u \in \mathcal{U}} ||\mathbf{w}_{u,b,q}||_2^2 z_{u, q} \leq  P_{b,q},\\
			 &\sum_{u \in \mathcal{U}} \tilde{R}_{u, q} \beta_{u,b,q} ||\mathbf{w}_{u,b,q}||_2^2 \leq C_{b, q}.
\end{align}  \label{opt_ch3:sum_ratio_problem}
\end{subequations}

The objective function, $f_r(\mathbf{w}, \mathbf{z}, \mathbf{\gamma})$, with a sum-of-ratio form, is expressed in  \eqref{eq: fr}. 

\begin{table*}
\centering
\begin{equation}
f_r(\mathbf{w}, \mathbf{z}, \mathbf{\gamma}) = \sum_{(u,q)}   \log_2 \left( 1 + \gamma_{u, q} \right) -  \gamma_{u, q}  + (1+\gamma_{u, q})\left(\frac{ z_{u,q}| \mathbf{h}_{u, q}^{\dagger}\mathbf{w}_{u,q}|^2}{\sigma^2  + \sum_{(i,j)} z_{i,j}|\mathbf{h}_{u,j}^{\dagger} \mathbf{w}_{i,j}|^2 } \right). \label{eq: fr}
\end{equation}
\medskip
\hrule
\end{table*}

\begin{table*}
\centering
 \begin{equation}
f_{FP}(\mathbf{w}, \mathbf{z}, \mathbf{\gamma}, \mathbf{y}) = \sum_{u,q}    \log_2 \left( 1 + \gamma_{u, q} \right) -   \gamma_{u, q} + 2z_{u,q}\sqrt{  (1 + \gamma_{u, q})}\text{Re}\{y_{u,q} \mathbf{w}_{u,q}^{\dagger}\mathbf{h}_{u, q} \} - |y_{u,q}|^2\sigma^2 - |y_{u,q}|^2\left(  \sum_{(i,j)} z_{i,j} \mathbf{h}_{u,j}^{\dagger}\mathbf{w}_{i,j}\mathbf{w}_{i,j}^{\dagger}\mathbf{h}_{u,j}  \right). \label{eq: FP_func}
\end{equation}
\medskip
\hrule
\end{table*}

The optimization problem in \eqref{opt_ch3:sum_ratio_problem}  can readily apply  quadratic transformation for multi-dimensional and complex FP \cite[Theorem 2]{FP1}, where the objective function is recasted as  in \eqref{eq: FP_func},
 where  $\mathbf{y} = \{y_{u,q} | (u, q) \in \mathcal{U} \times \mathcal{Q} \}$ is the vector of auxiliary variables, and its optimal value is found as

\begin{equation}
y_{u,q}^{\ast} = \frac{z_{u,q}| \mathbf{h}_{u, q}^{\dagger}\mathbf{w}_{u,q}|^2\sqrt{  (1 + \gamma_{u, q})}}{\sum_{(i,j)} z_{i,j}| \mathbf{h}_{u,j}^{\dagger}\mathbf{w}_{i,j}|^2}. \label{eq: y_star}
\end{equation}

Finally,  problem \eqref{opt_ch3:sum_ratio_problem} can be expressed as 
\begin{subequations}
\begin{align}
\max_{\mathbf{w}, \mathbf{z}, \mathbf{\gamma}, \mathbf{y}} \quad & f_{FP}(\mathbf{w}, \mathbf{z}, \mathbf{\gamma}, \mathbf{y})    \\
\textrm{s.t.}&~\sum_{q \in  \mathcal{Q}} z_{u, q} \leq 1,\\
			 &\sum_{u \in \mathcal{U}} ||\mathbf{w}_{u,b,q}||_2^2 z_{u, q} \leq  P_{b,q},\\
			 &\sum_{u \in \mathcal{U}_q} \tilde{R}_{u, q} \beta_{u,b,q} ||\mathbf{w}_{u,b,q}||_2^2 \leq C_{b, q},
\end{align} \label{eq: FP1}
\end{subequations}

\noindent where $ \mathcal{U}_q = \{ u \in \mathcal{U} | z_{u,q} = 1\}$. 

We note that fixing the auxiliary variable $\mathbf{y}$ results in $f_{FP}(\mathbf{w}, \mathbf{z}, \mathbf{\gamma}, \mathbf{y})$ being concave for the beamforming vector. On the other hand, if all other variables are fixed, $f_{FP}(\mathbf{w}, \mathbf{z}, \mathbf{\gamma}, \mathbf{y})$ is linear in $\mathbf{z}$. Thus, (\ref{eq: FP1}) can be optimized iteratively. The iterative algorithm that updates the variables must ensure that a particular association in one iteration can be updated in future iterations if needed, which is not the case when the iterative algorithm updates the variables independently. Therefore, we introduce new auxiliary variables for the beamforming vector by solving $\frac{\partial  f_{FP}}{\mathbf{w}_{u,q}} = 0~$ for $\mathbf{w}_{u,q}$, yielding the optimal auxiliary beamforming vector
\begin{equation}
\mathbf{v}_{u,b,q} = \frac{\sqrt{  (1 + \gamma_{u, q})}y_{u,q}\mathbf{h}_{u,b, q}}{\sum_{(i,j)} ||y_{i,j} \mathbf{h}_{i,q}||_2^2 + \eta_{b, q}^{\ast}}, \forall b \in \mathcal{B}_q, 
\end{equation}

\noindent where $\mathbf{h}_{u,b, q} \in \mathbb{C}^M$, and $\eta_{q, b}^{\ast}$ is a coefficient satisfying 
\begin{equation}
\eta_{b, q}^{\ast} =  \text{min} \{\eta_{b, q} \geq 0 \vert \sum_{u \in \mathcal{U}} ||\mathbf{v}_{u,b,q}||_2^2 \leq P_{b,q}  \}. 
\end{equation}

Next, we define a variable reflecting the benefits of assigning user $u$ to cluster $q$ as 
\begin{equation}
\begin{split}
\alpha_{u, q} &=  \log_2 \left( 1 + \gamma_{u, q} \right) -  \gamma_{u, q}
- |y_{u,q}|^2\sigma^2  - \sum_{(i,j)} |y_{i, j} \mathbf{h}_{i,q}^{\dagger}\mathbf{v}_{u,q}|^2 \\&+ 2\sqrt{  (1 + \gamma_{u, q})}\text{Re}\{y_{u,q} \mathbf{h}_{u,q}^{\dagger}\mathbf{v}_{u, q} \}, 
\end{split} \label{eq: alpha}
\end{equation}

\noindent where $\mathbf{v}_{u, q} = \{\mathbf{v}_{u,b,q} | (u, q) \in \mathcal{U} \times \mathcal{Q},  b \in \mathcal{B}_q \}$.

Thus, maximizing $f_{FP}$ in $\mathbf{z}$ is simply 
\begin{equation}
     z_{u,q} =
    \begin{cases}
      1, & \text{if}\ q = \underset{t \in \mathcal{Q}}{\mathrm{argmax}} ~ \alpha_{u, t} \\
      0, & \text{otherwise}. 
    \end{cases}\label{eq: Z_star}
\end{equation} 

Next, by fixing the optimzed association variables found in \ref{eq: Z_star}, we solve for the optimal beamforming vectors. That is 
\begin{subequations}
\begin{align}
\max_{\mathbf{w}} \quad & f_{FP}(\mathbf{w}, \mathbf{z}, \mathbf{\gamma}, \mathbf{y})    \\
\textrm{s.t.}&~\sum_{u \in \mathcal{U}} ||\mathbf{w}_{u,b,q}||_2^2 z_{u, q} \leq  P_{b,q},\\
			 &\sum_{u \in \mathcal{U}_q} \tilde{R}_{u, q} \beta_{u,b,q} ||\mathbf{w}_{u,b,q}||_2^2 \leq C_{b, q}. 
\end{align} \label{eq: FP2}
\end{subequations}

Problem (\ref{eq: FP2}) is a convex optimization problem since it maximizes a concave objective function in $\mathbf{w}$ and quadratic constraints, and be solved efficiently using CVX \cite{cvx}. Next, we describe the proposed iterative optimization algorithm in Section \ref{sec:sumrate_alg},  discuss its complexity performance in Section \ref{sec:sumrate_complx} and illustrate its performance using numerical results in Section \ref{sec:results}.

\subsection{Overall Algorithm}\label{sec:sumrate_alg}

Based on  Section \ref{sec:sumrate_reform} and \ref{sec:sumrate_sol} above, this paper proposes solving the optimization problem in \eqref{opt_ch3:Opt} using the iterative Algorithm \ref{alg:ch3}. Specifically,  the structure of \eqref{eq: FP_func} allows us to optimize over the optimization variables (i.e., beamforming vector and user association) iteratively. By doing so, we end up with an optimization problem over the binary variable only (line \ref{alg_line:z} of Algorithm \ref{alg:ch3}) and another optimization problem in the beamforming vectors (line \ref{alg_line:w} of Algorithm \ref{alg:ch3}). Hence, Algorithm \ref{alg:ch3} first activates a set of users that maximizes the quantity defined in \eqref{eq: alpha}. Algorithm \ref{alg:ch3} then fixes user-association $\mathbf{z}$ and optimizes over the associated beamforming vector $\mathbf{w}$ to maximize the network sum-rate. Additionally, we note that the two optimization steps in Algorithm \ref{alg:ch3} feature non-decreasing functions, which guarantees the algorithm convergence to a local optima, since the channel capacity upper-bounds the sum-rate, as the numerical simulations of Section \ref{sec:results} later show.

\begin{algorithm} 
\caption{User Clustering and Beamforming Iterative Algorithm}\label{alg:ch3}
\begin{algorithmic}[1]
\STATE Initialize $\mathbf{w}, \mathbf{z}, \mathbf{\gamma}$ to feasible values
\REPEAT
	\STATE update $\mathbf{\gamma}$ by (\ref{eq: gamma_star})
	\STATE update $\mathbf{y}$ by (\ref{eq: y_star})
 	\STATE optimize $\mathbf{z}$ by (\ref{eq: Z_star}) \label{alg_line:z}
 	\STATE solve (\ref{eq: FP2}) \label{alg_line:w}
	\STATE update $\tilde{\mathbf{R}}$ and $\mathbf{\beta}$
\UNTIL $f_{FP}$ converges
\end{algorithmic}
\end{algorithm}

\subsection{Complexity Analysis}\label{sec:sumrate_complx}
 The complexity of Algorithm \ref{alg:ch3} is dominated by the optimization problem solved numerically in step \ref{alg_line:w} of the algorithm, which is a convex quadratically constrained quadratic programming (QCQP) problem--a class of problems that are often expressed as second-order cone programming (SOCP)  \cite{lobo_socp}. 
 The SOCP problems are of great interest as they are featured in several practical engineering design problems. Several numerical solvers, such as MOSEK \cite{mosek}, can efficiently solve SOCP in polynomial time via interior-point methods \cite{nesterov_optim}.  

 In fact, in a worst-case performance, a complexity of the magnitude $\mathcal{O}(d^3)$ is required to solve a convex QCQP problem,  if $d$ is the size of the optimization variable \cite{luo_convex, ye_ipm}. Therefore, solving step \ref{alg_line:w} (i.e., (\ref{eq: FP2}))  requires a $\mathcal{O}(d^3)$ computational complexity, per path-following interior-point iteration, where $d$ is the number of variables considered in \eqref{eq: FP2}.

Despite the efficiency of Algorithm \ref{alg:ch3} in maximizing the network-wide throughput, where the numerical results of Algorithm \ref{alg:ch3} indeed later show the yielded gain in connecting the unconnected in a cloud-enabled HAPS system,  Section \ref{sec:results} also highlights that a significant fraction of the users remains unconnected. That is, the problem considered in this section deals with a linear function of the sum-rate. Hence, the proposed solution is a greedy solution that favors a subset of users with strong channel gain and those whose channels are nearly orthogonal (i.e., do not experience a large amount of interference). Although such a solution provides a high throughput, the solution tends to leave a percentage of users unserved, leading to a disparity in user rates. Motivated by the vision of 6G networks of democratizing access to connect the unconnected, the following section considers a utility function that provides a fair service solution. Section \ref{sec:sumlograte} aims to optimize an objective function that enforces high priority to serve low-rate users and the opposite for the privileged well-served users. To this end,  Section \ref{sec:sumlograte} deals with the logarithmic utility of the long-term average rate. Section \ref{sec:sumlograte} thoroughly discusses how the consideration of mentioned objective encourages load-balancing and accomplishes a proportional fairness service among users. 

\section{Load-Balancing via Beamforming and User Scheduling}\label{sec:sumlograte}

The previous section of this work considers a resource allocation approach that maximizes the network-wide sum-rate. Hence, the proposed solution tackles the problem by scheduling only a subset of users with favorable channel gains, leaving the remaining users unconnected. The numerical results featured later in Section \ref{sec:results} demonstrate Algorithm \ref{alg:ch3}'s potential in obtaining high sum-rates; however, this comes at the expense of increasing the disparity between user rates. Motivated by the goal of this article to connect the unconnected and exploit the powerful resources of HAPS-assisted systems to overcome digital inequality, this section focuses on devising a network objective that adopts fairness across users as an objective measure. In particular, this section investigates user scheduling policy and resource allocation techniques that used proportionally fair load-balancing as an objective.

The problem of scheduling users according to a fairness criterion has received a great deal of attention in the past literature. This article considers the proportional fairness criterion, which considers the aggregate utility/throughput and user fairness \cite{tse_pf}.
 In fact, in a network with scarce and limited resources, proportional fairness presents an elegant framework that achieves a trade-off between fairness and throughput \cite{bu_gpfs}, which has been widely used in practice \cite{tse_pf}. Moreover, the analysis and convergence of proportional fairness sharing algorithms have been studied in \cite{kushner, kelly_pfs}. Therefore, this work now considers a sum-log of the long-term average rate objective function, and derives an iterative-based weighted sum-rate maximization problem, as this section further elaborates.

\subsection{Problem Statement}

Consider the system model in Section \ref{sec:system_model},  consisting of a multilayered cloud-enabled HAPS network with $B$ BSs. 
Toward the goal of accommodating more users, this section adopts the separation of users in frequency via an OFDMA scheme and focuses on scheduling users to $N$ frequency tones (i.e., subcarriers). Such a scheme is a multiuser access scheme that divides the channel bandwidth into $N$ orthogonal narrow-band subcarriers/subchannels---where the subcarriers experience frequency selectivity. Furthermore, the OFDMA scheme exploits multiuser diversity gains by allocating different subcarriers to different users \cite{song_ofdm}. Unlike Section \ref{sec:sumrate}, which dynamically associates users to clusters, this section assumes fixed user clustering and focuses on user scheduling to frequency tones. The signal model presented in \ref{sec:system_model} is readily extendable to the OFDMA case. The received signal at user $u$ on the $n$-th subcarrier can be expressed as

\begin{equation}
y_{u,q}^n = (\mathbf{h}_{u,q}^n)^{\dagger}\mathbf{w}_{u,q}^n s_u  + \sum_{\substack{(i,j) \neq (u,q), \\ i \in \mathcal{U}, j \in \mathcal{Q}}} (\mathbf{h}_{u,j}^n)^{\dagger} \mathbf{w}_{i,j}^n s_i  + \nu_u^n, 
\label{eq:yn}
\end{equation}
\noindent where the superscript $n$ denotes the frequency tone. For example,  $\mathbf{h}_{u,q}^n \in \mathbb{C}^{ML \times 1}$ is the channel vector of user $u$ served by cluster $q$ in tone $n$. Let $S(q,b,n)$ be a set-valued function determining the scheduling policy of a subset of users $\mathcal{S}_{q,n}$ on the $n$-th tone of the $b$-th BS in cluster $q$. Moreover, let $\mathcal{T}_{u,q} = \{ n  | u \in \mathcal{S}_{q, n} \}$ be the set of tones assigned for user $u$. Then, the rate of user $u$ is 

\begin{equation}
R_{u, q} =  \sum_{n \in \mathcal{T}_{u,q}} \log_2 \left( 1 + \mathrm{SINR}_{u,q}^n \right), \label{eq: rate_ofdma}
\end{equation}
\noindent where the SINR of user $u$  served by the $q$-th cluster at the $n$-th subcarrier is
\begin{equation}
\mathrm{SINR}_{u,q}^n = \frac{| (\mathbf{h}_{u, q}^n)^{\dagger}\mathbf{w}_{u,q}^n|^2 }{ \sum_{(i,j) \neq (u, q)}|(\mathbf{h}_{u, j}^n)^{\dagger} \mathbf{w}_{i,j}^n|^2  + \sigma^2}.  \label{eq: SINR_n}
\end{equation}

As the previous section demonstrates, the sum-rate maximization problem focuses on users with strong and favorable channels, leaving other users idle. Instead, this section considers the maximization of a network utility function, typically an increasing concave function, where we consider the logarithm function and the utility as the long-term average rate of users. The choice of this objective function is common in wireless communication networks and has been shown to be optimal in creating a fair service \cite{load_andrews}. Indeed, the maximization of network utility reflects a load-balancing and fair scheduling framework because the increasing, concave utility function presents a diminishing return paradigm in which serving good/opportunistic users leads to a marginal increase in the network utility. On the contrary, serving those users with relatively poor  channel gains, which are often not selected, results in higher gains. Therefore, the max-sum-log-throughput framework rewards serving more users and encourages the scheduling policy to activate users which have not been scheduled before.

Motivated by the above discussion, this section considers the design of a user scheduling function $S(q,b,n)$  and the beamforming vector jointly, to maximize 

\begin{subequations}
\begin{align}
\max_{\mathbf{w},   \mathcal{S}, \mathbf{r}} \quad & \sum_{  (u, q) \in \mathcal{U} \times \mathcal{Q}} \log{ \left( \overline{R_{u, q}} \right)} \label{opt_ch4:obj_ofdma}  \\
\textrm{s.t.}&~ (\ref{eq: rate_ofdma}) \\
			 & \sum_{n \in \mathcal{T}_{u,q}}\sum_{u \in \mathcal{U}_q} ||\mathbf{w}_{u,b,q}^n||_2^2 \leq  P_{b,q},\label{opt_ch4:pwr_ofdma}\\
			 & \sum_{n \in \mathcal{T}_{u,q}} \sum_{u \in \mathcal{U}_q} R_{u, q}^n \mathbbm{1} \{||\mathbf{w}_{u,b, q}^n||_2^2\} \leq C_{b, q},\label{opt_ch4:cap_ofdma_indic}
\end{align}\label{opt_ch4:Opt}
    \end{subequations}
\noindent where $\overline{R_{u, q}}$ is the long-time averaged rate, updated exponentially over a duration of $T$ time slots, for some $\eta \in (0,1) $ as
\begin{equation}
	\overline{R_{u, q}}^t = (1-\eta) \overline{R_{u, q}}^{t-1} + \eta R_{u,q}^t, \label{eq: Ravg}
\end{equation}
\noindent where $R_{u,q}^t$ is the rate obtained by user $u$ in time slot $t \leq T$, as defined in \eqref{eq: rate_ofdma}. For simplicity, we emit the notation of  $t$ throughout the discussion. 

It is worth highlighting that the constraints presented in problem \eqref{opt_ch4:Opt} are across all tones of BS $b$. Hence, the problem cannot be decoupled on a per-tone basis, as done in \cite{yu_multicell}. Similar to \eqref{opt_ch3:Opt},  problem \eqref{opt_ch4:Opt} is a mixed discrete and continuous non-convex optimization problem. Problems of this nature have combinatorial complexity \cite{sched_combin} due to the binary variable of user scheduling. Therefore, this section employs an iterative approach to solving the problem, where the paper first solves the problem for user scheduling, then fixes the user scheduling and solves for the beamforming vectors subproblem. The proposed approach mimics a coordinate-ascent approach and can guarantee local optimality.

\subsection{Proposed Solution}\label{sec:sumlograte_proposedsol}

This section presents a system-level solution to problem \eqref{opt_ch4:Opt}. First, we optimize \eqref{opt_ch4:Opt} over the user scheduling, where we adopt a proportionally-fair criterion while holding the beamforming vectors fixed. Then, we use the solution for the scheduling policy and solve for the beamforming vector under physical resource constraints. 

\subsubsection{User Scheduling}\label{sec:sumlograte_sched}

The problem of obtaining an optimal scheduling policy is combinatorial and intractable. Hence,  in this section, we seek to find a locally optimal solution that can incrementally maximize the objective function \eqref{opt_ch4:obj_ofdma} \cite{tse_pf}. Solving the optimization problem iteratively is motivated by observing the interference pattern created in the downlink direction, where the power leakage of a signal from  BS $b$ to user $u$ is fixed, is independent of the user assignment to the serving subcarrier. In other words, it is possible to decouple the problem and tackle it using coordinate ascent since the mapping/assignment of users to frequency tones is independent of the beamforming vector.

The user scheduling problem boils down to 
\begin{subequations}
\begin{align}
\max_{  \mathcal{S}, \mathbf{r}}  \quad  &\sum_{  (u, q) } \log{ \left( \overline{R_{u, q}} \right)}, \label{opt_ch4:sched}\\
\textrm{s.t.}&~ (\ref{eq: rate_ofdma}), \eqref{opt_ch4:pwr_ofdma}, \eqref{opt_ch4:cap_ofdma_indic},
\end{align} 
\end{subequations}
\noindent which can be easily solved while ensuring proportional fair scheduling. It is easy to see that the maximization of sum-log-throughput can be recast as a weighted sum-rate maximization, with weights adaptively selected to be proportional to the reciprocal incrementally updated average rate. In other words, note that  
\begin{equation} \label{eq:computeWeight}
	 \left.\frac{\partial \log(\overline{R_{u, q}}^t)}{\partial R}\right|_{\overline{R_{u, q}}^t}
        = \frac{1}{\overline{R_{u, q}}^t}.
\end{equation}

Let $\alpha_u = \frac{1}{\overline{R_{u, q}}^t}$, then \eqref{opt_ch4:sched} translates to 
\begin{equation}
\max  \quad  \sum_{  (u, q) } \alpha_u R_{u, q}.
\end{equation} \label{opt_ch4:wsr}

Thus, we conclude that the contribution of the achievable rates over a window $T$ is equivalent to a weighted sum-rate \cite{berry, kushner}. Finally, a scheduling policy that obeys the proportional fairness criterion is expressed as 
\begin{equation}
    S(q,b,n) = \underset{u \in \mathcal{U}_q}{\mathrm{argmax}} ~ \frac{R_{u, q}^n}{\overline{R_{u, q}}}. \label{eq: scheduling}
\end{equation}

\subsubsection{Beamforming}\label{sec:sumlograte_beamf}

Following the discussion of user scheduling optimization, optimizing the beamforming vectors for a fixed user assignment corresponds to maximizing the weighted sum-rate.
 Therefore, the objective is to solve the following 
\begin{subequations}
\begin{align}
\max_{\mathbf{w}, \mathbf{r}} \quad & \sum_{  (u, q) \in \mathcal{U} \times \mathcal{Q}}  \alpha_u R_{u, q}   \\
\textrm{s.t.}&~ (\ref{eq: rate_ofdma}), \eqref{opt_ch4:pwr_ofdma}, \eqref{opt_ch4:cap_ofdma_indic},
\end{align}\label{opt_ch4:bmfrmn_prob}
\end{subequations}
\noindent where $\alpha_u  = \overline{R_{u, q}}^{-1}$ is the weight associated with user $u$, which determines its priority to be served based on the proportional fairness criterion \eqref{eq:computeWeight}. Problem \eqref{opt_ch4:bmfrmn_prob} has a similar structure to \eqref{opt_ch3:Opt}, except that it does not involve optimization over the discrete variable $\mathbf{z}$. Hence, the solution proposed to solve \eqref{opt_ch4:bmfrmn_prob} follows a similar approach to the one adopted in Section \ref{sec:sumrate_reform} and \ref{sec:sumrate_sol} in handling the indicator function and solving the non-convex problem.

Specifically, by approximating the indicator function \eqref{opt_ch4:cap_ofdma_indic} using the $\ell_0$-norm, and again with a re-weighted $\ell_1$-norm, as outlined in  \eqref{eq:ell0}, we arrive to the following problem

\begin{subequations}
\begin{align}
\max_{\mathbf{w}, \mathbf{r}} \quad &\sum_{  (u, q) \in \mathcal{U} \times \mathcal{Q}} \alpha_u R_{u, q} \label{eq: opt_w_objective}  \\
\textrm{s.t.}&~ (\ref{eq: rate_ofdma}), (\ref{opt_ch4:pwr_ofdma}), \\
			 & \sum_{n \in \mathcal{T}_{u,q}} \sum_{u \in \mathcal{U}_q}  \tilde{R}_{u, q}^n\beta_{u,b,q}^n ||\mathbf{w}_{u,b,q}^n||_2^2  \leq C_{b, q}, \label{eq_opt: cap_ofdma}
\end{align} \label{eq: opt_w}
\end{subequations}
\noindent where the approximation weights $\beta$ is defined as 
\begin{equation}
\beta_{u,b,q}^n= \frac{1}{\epsilon + ||\mathbf{w}_{u,b,q}^n||_2^2}. \label{eq: beta}
\end{equation}

Problem \eqref{eq: opt_w} remains a non-convex optimization problem because it features a non-convex objective function. We adopt a quadratic transform FP-based solution, where we find a direct solution using \cite[Theorem 3]{FP1} for a sum-of-functions-of-ratio similar to Section \ref{sec:sumrate_sol}.

Applying the quadratic transform for multidimensional and complex ratios theorem to each SINR term in \eqref{eq: opt_w_objective} allows recasting the problem as 
\begin{subequations}
\begin{align}
\max_{\mathbf{w}} \quad & f_{FP}(\mathbf{w}, \mathbf{y})    \\
\textrm{s.t.}&~(\ref{opt_ch4:pwr_ofdma}), \eqref{eq_opt: cap_ofdma}
\end{align} \label{eq: FP2_ofdma}
\end{subequations}
\noindent where

\begin{equation}
\begin{split}
f_{FP}(\mathbf{w}, \mathbf{y}) = & \sum_{(u,q,n)}\alpha_u \log \Biggl(  1 + 2\text{Re}\{y_{u,q}^n (\mathbf{h}_{u,q}^n)^{\dagger}\mathbf{w}_{u, q}^n \} \\&- |y_{u,q}^n|^2\left( \sigma^2 +  \sum_{(i,j) \neq (u,q)} |(\mathbf{h}_{u,j}^n)^{\dagger}\mathbf{w}_{i,j}^n|^2  \right) \Biggr),
\end{split}\label{eq: f_cf}
\end{equation}
\noindent and the optimal value of the auxiliary variable, determined by solving $\frac{\partial  f_{FP}}{\mathbf{w}_{u,q}} = 0$, is

\begin{equation}
(y_{u,q}^n)^{\ast} = \frac{ (\mathbf{h}_{u, q}^n)^{\dagger}\mathbf{w}_{u,q}^n  }{\sum_{(i,j)\neq (u,q)} \sigma^2 + | (\mathbf{h}_{u,j}^n)^{\dagger}\mathbf{w}_{i,j}^n|^2}. \label{eq: y_CF}
\end{equation}

Note that the quadratic transform in this section decouples and recasts the objective function to a sum-of-log. This is in contrast to the quadratic transform employed in Section \ref{sec:sumrate}, which rather recasts the objective function to sum-of-ratios and avoids the logarithm function. In fact, a similar approach to the one featured in Section \ref{sec:sumrate} can be used here by simply formulating the Lagrangian dual of \eqref{eq: opt_w}, which offers a closed-form solution. Nonetheless, the problem offered by the direct quadratic transform approach proposed in our paper yields a convex optimization in $\mathbf{w}$ as seen from \eqref{eq: FP2_ofdma}, since function \eqref{eq: f_cf} is concave.

\subsection{Overall Algorithm}

The optimization problem featured in this section is a combinatorial non-convex mixed integer-continuous optimization problem. This paper adopts an iterative approach to tackle the intricacies of \eqref{opt_ch4:Opt} based on the discussion in Section \ref{sec:sumlograte_proposedsol}.  The proposed algorithm features two nested loops, one for optimizing the scheduling policy and the other for optimizing the beamforming vectors. More specifically, in the outer loop, the algorithm determines user scheduling according to the proportional fairness criterion \eqref{eq: scheduling} and updates the priority weights of the users. On the other hand, by fixing the weights, the algorithm solves the weighted sum-rate optimization problem over the associated beamforming vectors \eqref{eq: FP2_ofdma} in the inner loop. Moreover, the weighted sum-rate optimization problem is guaranteed to converge since $f_{FP}(\mathbf{w}, \mathbf{y})$ is a non-decreasing and upper-bounded function of the rate. Algorithm \ref{alg:ch4} summarizes the steps of the numerical optimization algorithm solving \eqref{opt_ch4:Opt}.

\subsection{Complexity Analysis}
Similar to the discussion in Section \ref{sec:sumrate_complx}, solving \eqref{eq: FP2_ofdma} numerically creates the highest complexity in the proposed algorithm. The problem in \eqref{eq: FP2_ofdma} is also a convex QCQP that can be remodeled as a SOCP and solved using numerical optimization solvers such as those provided by CVX \cite{cvx}. 
Therefore,  the worst-case performance optimization for SOCP depends on the size of the optimization variables in problem \eqref{eq: FP2_ofdma} $d$, and given by $\mathcal{O}(d^3)$.

\begin{algorithm}
\caption{User Scheduling and Beamforming Iterative Algorithm}\label{alg:ch4}
\begin{algorithmic}
\STATE Initialize $\mathbf{w},  \mathcal{S}$ to feasible values
\REPEAT 
\STATE update $\overline{R}$ using (\ref{eq: Ravg})
\STATE update scheduling using (\ref{eq: scheduling}) and the weights using (\ref{eq:computeWeight})
\STATE fix $\tilde{R}$, and update $\beta$ using (\ref{eq: beta})
    \REPEAT
    	\STATE update $\mathbf{y}$ by (\ref{eq: y_CF})
     	\STATE solve (\ref{eq: FP2_ofdma}) 
    \UNTIL $f_{FP}$ converges
\UNTIL  $T$ iterations
\end{algorithmic}
\end{algorithm}

\section{Numerical Results}\label{sec:results}

This section aims to study the effectiveness of the resource management solutions proposed for cloud-enabled HAPS systems. Specifically, we investigate the performance of Algorithm \ref{alg:ch3} and Algorithm \ref{alg:ch4} in achieving a high sum-rate and providing an equitable service framework, respectively. This section starts by evaluating the performance of iterative user clustering and beamforming techniques in Algorithm \ref{alg:ch3} against classical approaches for beamforming and user clustering, as well as a standalone terrestrial communication system. The numerical results in this section present a trade-off in the coordination among BSs in clusters and provide a visualization of the gain obtained from adopting HAPS-assisted systems serving more users in low-density and unpopulated areas. This section then shows the shortcomings of adopting a throughput maximization scheme in leaving users with low channel gains unserved. Hence, the section provides a numerical example that motivates the problem featured in Section \ref{sec:sumlograte}, which presents a load-balancing framework that can accomplish a proportional fairness service among users. Similarly, the section studies the effectiveness of Algorithm \ref{alg:ch4} in a cloud-enabled HAPS system, investigates the performance of the proposed user assignment and beamforming algorithm in an OFDMA case, and compares it to a single-carrier system, such as the one featured in Section \ref{sec:sumrate}, and a standalone terrestrial communication system. 

\subsection{Throughput Maximization}
We consider a square area of  $300~ \mathrm{km}^2$ and simulate various density areas (e.g., urban, suburban, and rural areas). The network set-up consists of $U = 100$ users, where $U_A = 30,~ U_T = 70$, and $B = 48$, where $B_T = 18$ and $B_A = 30$ BSs. Furthermore, we assume the channel between  TBSs and terrestrial users includes (i) the deterministic path loss component, (ii) a large-scale log-normal fading that characterizes the shadowing phenomena, (iii) and small-scale Rayleigh fading that characterizes the scattering phenomena. On the other hand, the channel of the users served by  HBSs includes (i) the deterministic path loss component and (ii) a small-scale Rician fading \cite{alsharoa}. Table \ref{table:system-parameter} lists the simulation parameters used in this section. Moreover, we obtain the simulation results using CVX \cite{cvx}; specifically the MOSEK solver \cite{mosek}.

\begin{table}
\centering
\caption{Simulation parameters}
\label{table:system-parameter}
\begin{tabular}{|l|l|}
\hline 
Parameter  & Value  \\\hline
Bandwidth & $10$ MHz    \\ \hline
Max. Tx power $P$  &  $32$ dBm \\ \hline
Max. fronthaul capacity $C$  &  $500$ Mbps \\ \hline
Num. of BS antenna $M$  &   $2$ \\ \hline
 Noise power spectral density  & $-169$ dBm/Hz \\ \hline
 Max. altitude of HBS & $2$ km \\ \hline
 Max. altitude of UAVs & $100$ m \\ \hline
 Path loss exponent for TBS  & $ 3.76$ \\ \hline
 Path loss exponent for HBS & $2$ \\ \hline
 Shadowing standard deviation & $8$ dB \\ \hline
 Rayleigh fading power & $0$ dB  \\ \hline
 Rician factor & $10$ dB  \\ \hline
 Regularization  parameter $\epsilon $ & $ 10^{-12}$ \\ \hline
\end{tabular}
\end{table}

\begin{figure}
    \centering
    \includegraphics[width=\linewidth]{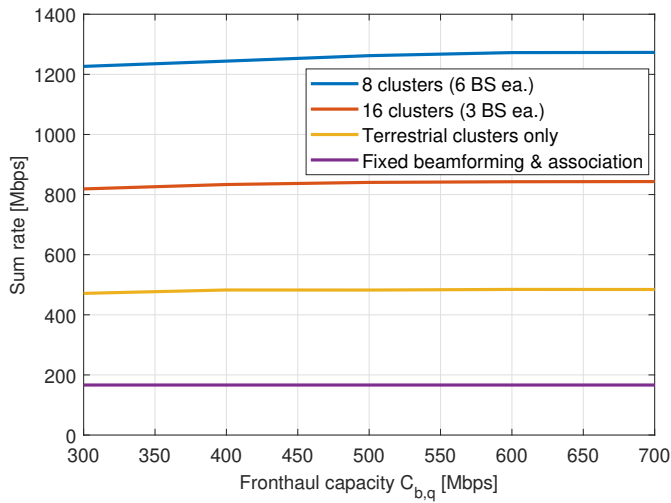} 
    \caption{Gain achieved by using two tiers compared to a single tier in a cloud-enabled HAPS system versus the available fronthaul capacity.}\label{fig:ch3_capacity}
\end{figure}

One of the important metrics to assess is the performance of the optimization algorithm subject to resource availability. Hence, we start by evaluating the sum-rate yielded by Algorithm \ref{alg:ch3}  when changing the resource availability. Fig.~\ref{fig:ch3_capacity} shows the sum-rate gain for different values of available fronthaul capacity. In particular, we assess the proposed algorithm using three scenarios. First, we show eight clusters, each with six BSs; three of these clusters are terrestrial clusters. Second, we spread the BSs across 16 clusters (six terrestrial clusters and ten hot-air balloon clusters), each with three BSs. Then, we simulate the proposed algorithm on a standalone terrestrial system with three terrestrial clusters. We analyze the performance of the proposed dynamic clustering and beamforming algorithm with fixed clustering and beamforming methods (i.e., user clustering according to the shortest average distance and beamforming using the zero-forcing technique).

The results in Fig.~\ref{fig:ch3_capacity} show the sum-rate improvement when complementing a terrestrial communication system with a cloud-enabled HAPS and  HBSs. Equipping the terrestrial standalone communication system with HBSs, all connected to the HAPS cloud, yields at least a 70\% higher sum-rate. 
Another important result highlighted in Fig.~\ref{fig:ch3_capacity} is the impact of clustering BSs in coordinating BSs to serve users. Notably, forming the available $B$ BSs into eight clusters yields a higher sum-rate than forming them into 16 clusters. This result suggests that forming a cluster with a larger number of participating BSs allows for a higher level of coordination and fine-grain interference management between the BSs. The high coordination between BSs comes at the cost of higher complexity. Although quantifying such complexity is outside the scope of this study, the above result serves to provide a qualitative description of the computation versus communication interplay at the cloud level of the cloud-enabled HAPS.

\begin{figure}
    \centering
    \includegraphics[width=\linewidth]{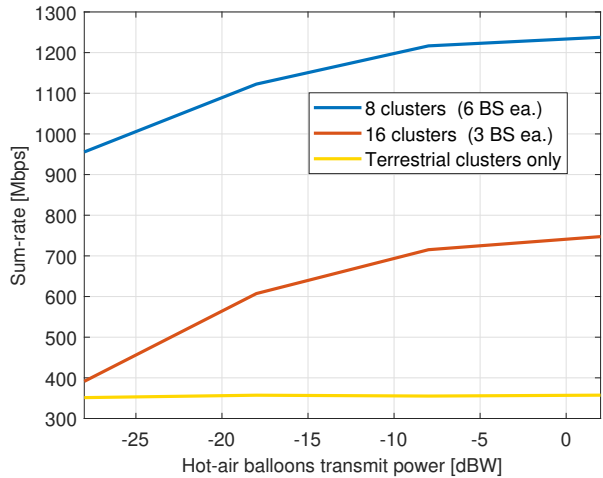} 
    \caption{Gain achieved by using two tiers compared to a single tier in a cloud-enabled HAPS system, as a function of the maximum transmit power at the HBSs.}\label{fig:ch3_power}
\end{figure}

Fig.~\ref{fig:ch3_capacity} compares the proposed dynamic user clustering and beamforming solution with fixed user clustering and beamforming techniques. As the figure illustrates, adopting the proposed solution provides three times the sum-rate performance given the same number of BSs in a HAPS-assisted system. Moreover, it is evident from Fig.~\ref{fig:ch3_capacity} that the proposed user clustering and beamforming technique in a standalone terrestrial system performs better than a HAPS-assisted system with fixed clustering and beamforming.

Next, Fig.~\ref{fig:ch3_power} shows the sum-rate gain of the proposed algorithm in a cloud-enabled HAPS as a function of the HBSs' available transmit power. Specifically, Fig. \ref{fig:ch3_power} demonstrates the resilience and robustness of the proposed cloud-enabled HAPS system under limited transmit power for the battery-powered  HBSs. That is, since HBSs are expected to serve users until deflated or out of battery, Fig. \ref{fig:ch3_power} shows that the proposed system model can still achieve gains in terms of network sum-rate even when the HBSs are close to being deflated. Moreover,  Fig. \ref{fig:ch3_power} highlights the significant sum-rate gains in spreading the BSs into a low number of clusters (e.g., 8 clusters instead of 16). Such gains come at the cost of higher computation complexity to coordinate the BSs within a cluster.

     \begin{figure}
      \centering
\includegraphics[width=\linewidth]{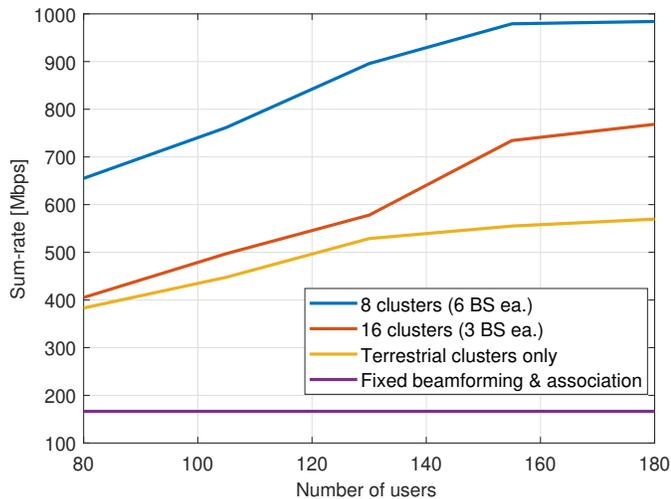} 
\caption{Gain achieved by using two tiers compared to a single tier, as the number of users grows. \label{fig: rateVuser}}
    \end{figure}

 Moreover, we show the improvement of the sum-rate versus the density of users in Fig.~\ref{fig: rateVuser}. Increasing user density allows the BSs to have better candidates, which increases the sum-rate objective function. Fig.~\ref{fig: rateVuser} also illustrates how the gain conferred by increasing the user density saturates after a certain threshold, beyond which dropping more users does not inflict further increase in sum-rate, as neighboring users would have similar channel conditions.

\begin{figure}
\centering
\subfloat[]{\includegraphics[width=\linewidth]{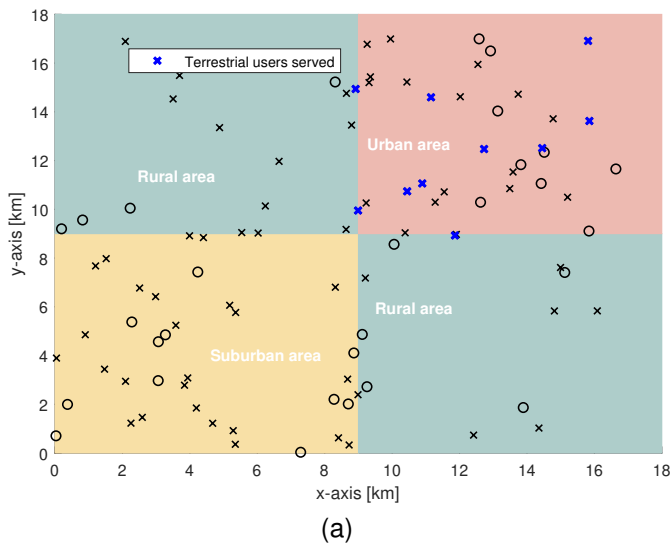}%
\label{fig:served_T}}
\hfil
\subfloat[]{\includegraphics[width=\linewidth]{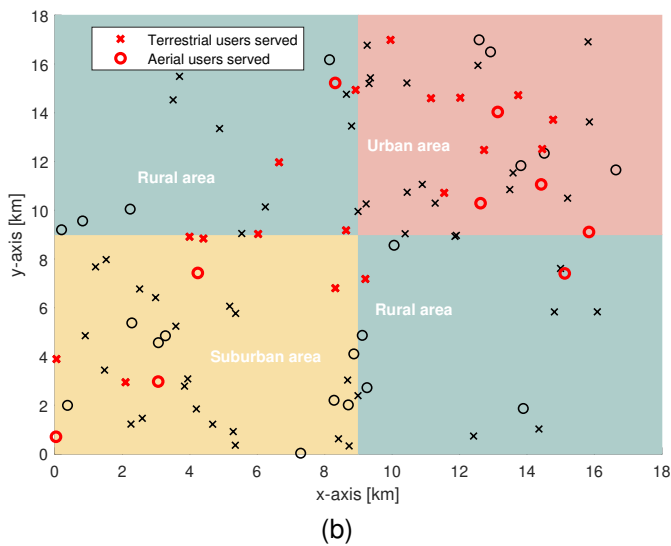}%
\label{fig:served_AG}}
\caption{Visualization of served users in a cloud-enabled HAPS system. (a) Users served by a terrestrial standalone system. (b) Users served by hot-air balloons and terrestrial clusters.}
\label{fig:served_users}
\end{figure}

We next show the gain obtained from a cloud-enabled HAPS system compared to a standalone terrestrial communication system in connecting the unconnected while ensuring coverage and capacity for users in hyper-digitized areas. In Fig.~\ref{fig:served_users}, we visualize the served aerial and terrestrial users in different geographical areas for one instance of user deployment in the network. Fig.~\ref{fig:served_T} highlights the terrestrial users served in a standalone terrestrial system. The served users are clustered in the urban area, while users in the suburban and rural areas are unconnected. Augmenting terrestrial communication systems with  HBSs in a cloud-enabled HAPS system allows more users to be served, as Fig.~\ref{fig:served_AG} illustrates. Particularly, a HAPS-assisted system offers coverage for users in low-density and unpopulated areas while ensuring connected areas continue receiving service. 
While HAPS-assisted systems can connect more users, Fig.~\ref{fig:served_users} shows that a sum-rate maximization solution serves around 30\% of the users. This section's sum-rate maximization objective yields a greedy solution that chooses a subset of users whose channel gain is favorable and leaves other users unserved.

\begin{figure}
    \centering
    \includegraphics[width=\linewidth]{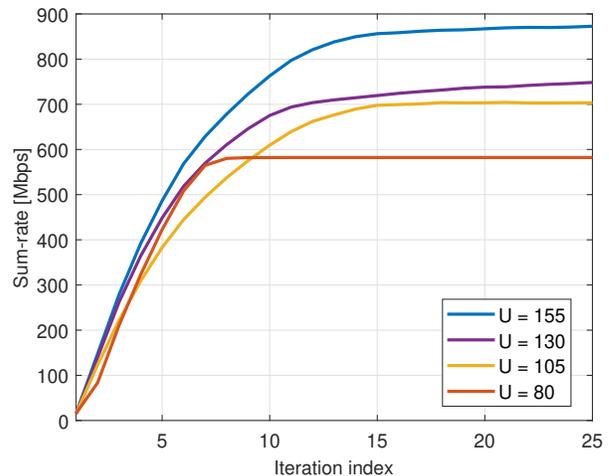} 
    \caption{Convergence of the sum-rate in Algorithm \ref{alg:ch3}. \label{fig: ch3converge}}
\end{figure}

Before concluding this part, we illustrate the convergence behavior of Algorithm \ref{alg:ch3} in Fig.~\ref{fig: ch3converge} by showing the sum-rate evolution over the number of iterations. Fig.~\ref{fig: ch3converge} illustrates the relatively fast convergence behavior of Algorithm \ref{alg:ch3}. Each iteration shown in Fig.~\ref{fig: ch3converge} represents the joint optimization over the association and beamformers, as well as the adaptive weights of the approximated $\ell_0$-norm (i.e., lines 3-7 of  Algorithm \ref{alg:ch3}). 
Fig.~\ref{fig: ch3converge} further highlights the gain in sum-rate obtained by implementing Algorithm \ref{alg:ch3}, where the algorithm shows around a  ten-fold improvement compared to the initial value in only ten iterations.


\subsection{Load Balancing}

\begin{figure}
    \centering
    \includegraphics[width=\linewidth]{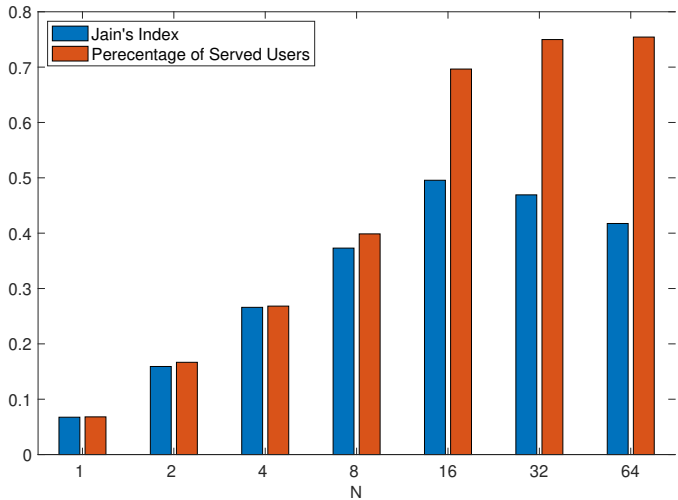} 
    \caption{Performance of a standalone terrestrial communication system of 36 users and 6 TBSs.}\label{fig:terr_ofdma}
\end{figure}

This section studies the effectiveness of Algorithm \ref{alg:ch4} in a cloud-enabled HAPS system for load-balancing purposes.  The proposed solution in Algorithm \ref{alg:ch4} employs two approaches different from those proposed in Algorithm \ref{alg:ch3}: considering an OFDMA scheme and maximizing a sum-log-throughput. Hence, this article evaluates the effectiveness of these two approaches in accommodating more users and promoting digital inclusion and equity framework in cloud-enabled HAPS systems. Having said that, first, we show the performance of Algorithm \ref{alg:ch4} in a standalone terrestrial communication system in Fig.~\ref{fig:terr_ofdma}. We consider $B = B_T = 6$ BSs, and $U = U_T = 36$ users. Particularly, we test the system's performance as we vary $N$, where $N = 1$ refers to a single carrier system, similar to Section \ref{sec:sumrate}. We evaluate the fairness among user rates using Jain's fairness index, $J \in [\frac{1}{U}, 1]$, \cite{jain}, which can be expressed as 
\begin{equation}
    J = \frac{(\sum_{\mathcal{U}} R_i)^2}{U \sum_{\mathcal{U}} R_i^2}. \label{eq:jain}
\end{equation}

We illustrate Jain's fairness index  \eqref{eq:jain} and the percentage of served users in a terrestrial-only communication system in Fig.~\ref{fig:terr_ofdma}. As the figure illustrates, a single-carrier system displays a high disparity in serving users, where no more than 10\% of the users are served. Such disparity in users' service motivates adopting an OFDMA scheme, where a simple OFDMA system improves user fairness by almost 40\% more than a single carrier system. Moreover, Fig.~\ref{fig:terr_ofdma} shows that an OFDMA system of 16 subcarriers/resource blocks yields six-fold the fairness of a single-carrier system. In fact,  such an increase in fairness is clear through the shown percentage of served users, which goes up to 70\%.

Interestingly,  Fig.~\ref{fig:terr_ofdma} shows that increasing the number of subcarriers beyond 16 in this scenario decreases the fairness index, while the number of served users increases but saturates eventually. While this observation is somewhat counter-intuitive, we note that it is still consistent with our original claim for the following reasons.
First, Algorithm \ref{alg:ch4} aims to provide an opportunity to serve more users by penalizing over-serving the users with strong channel gains and rewards serving users with mediocre channel gains. This, however,  does not mean that Algorithm \ref{alg:ch4} encourages draining its available resources to serve users with extremely poor channels. The rationale behind such observation is that serving users with extremely poor channels might result in dedicating most of the resources to serve the users with poor channels, and users with  better  channels would remain idle. Therefore,  Algorithm \ref{alg:ch4} gracefully and intelligently allocates the available resources to increase fairness  without suffocating  the sum-rate in the process. Second,  recall that Algorithm \ref{alg:ch4} promotes a fair service by penalizing over-serving users with strong channels to calibrate the rates. After a certain point, the number of served users saturates, and Algorithm \ref{alg:ch4} allocates the resources among the users in the active set. By doing so, the fairness index drops because 15\% of the users remain unserved, and allocating resources to  better  users increases the sum-rate. Noticeably, however, such a drop in the fairness index is insignificant compared to that of a  single carrier system, which highlights the dual numerical benefit of the proposed scheme (i.e., improving fairness for a negligible drop in rate).

\begin{figure}
    \centering
    \includegraphics[width=\linewidth]{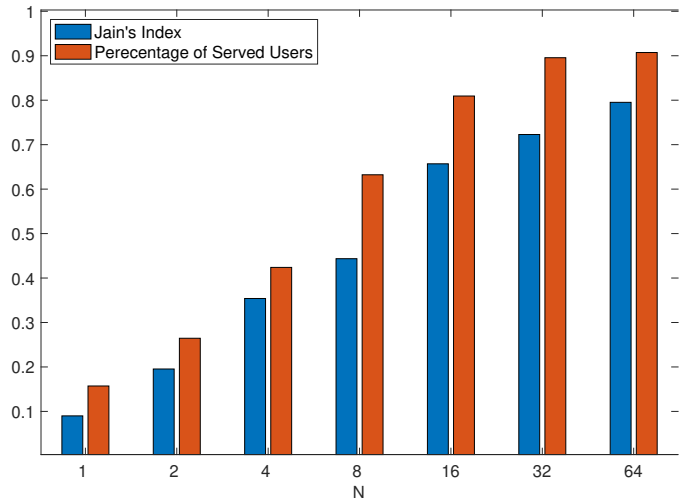} 
    \caption{Performance of a cloud-enabled HAPS system of 36 users, 2 clusters, each 3 BSs.}\label{fig:AT_ofdma}
\end{figure}

Next, we evaluate the performance of Algorithm \ref{alg:ch4} in a cloud-enabled HAPS system that integrates  TBSs and HBSs . We illustrate the performance of the proposed algorithm in Fig.~\ref{fig:AT_ofdma}. Similar to the standalone system in Fig.~\ref{fig:terr_ofdma}, adopting an OFDMA scheme, even for the simple case of $N = 2$, improves the fairness measure. Most importantly, in this scenario, Jain's fairness index is non-decreasing and offers up to eight times more fairness than a single carrier system. In fact, a cloud-enabled HAPS system, equipped with  both HBSs and TBSs, can serve up to 90\% of the users, while the standalone terrestrial system stops at 75\%, which offers another numerical testimony of the prospects of the proposed non-terrestrial network scheme in boosting the digital equity of the wireless systems under study.

\begin{figure}
\centering
\subfloat[]{\includegraphics[width=\linewidth]{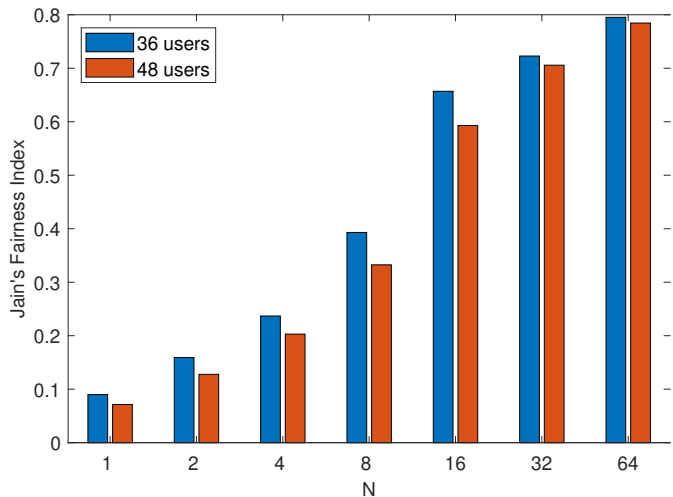} %
\label{fig:jain_2scen}}
\hfil
\subfloat[]{\includegraphics[width=\linewidth]{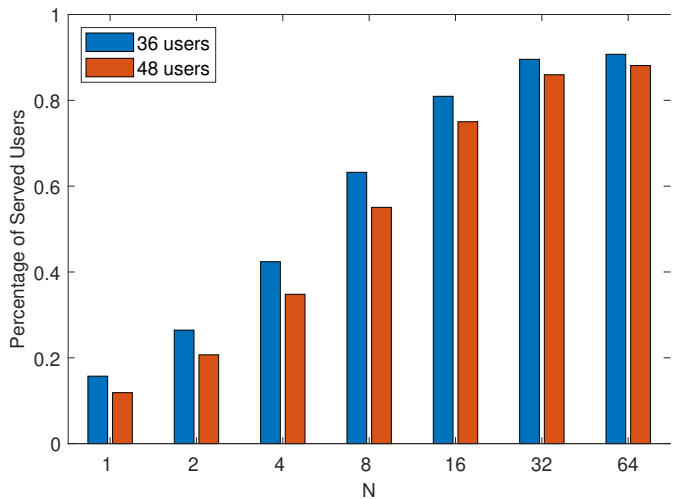} %
 \label{fig:users_2scen}}
\caption{Comparison of two scenarios of cloud-enabled HAPS systems. (a) Fairness measure of a  cloud-enabled HAPS system as the number of users increases. (b) Percentage of served users in a  cloud-enabled HAPS system as the number of users increases.}
\label{fig:ofdma_2scen}
\end{figure}

Comparing the fairness patterns obtained in Fig.~\ref{fig:terr_ofdma} and Fig.~\ref{fig:AT_ofdma} supports our claims that the proposed cloud-enabled HAPS system provides a ubiquitous connectivity framework to connect the unconnected while ensuring continuous service for the connected. In Fig.~\ref{fig:terr_ofdma}, the fairness measure drops because some users have poor channel conditions, and serving them using the available terrestrial infrastructure means one must sacrifice the sum-rate and starve users with good channels. On the other hand, Fig.~\ref{fig:AT_ofdma} shows that those users who are previously unserved due to their poor channel conditions can have a better channel with  HBSs. That is, since the channel between  HBSs  and users is a Rician fading channel with no shadowing component, as opposed to the Rayleigh channel of TBSs, the  HBSs  can better serve the users that have poor channels with  TBSs. This conclusion is consistent with our motivation to adopt  an OFDMA-based cloud-enabled HAPS system using an optimized proportional fairness scheduling and beamforming scheme.

\begin{figure}
\centering
\subfloat[]{\includegraphics[width=\linewidth]{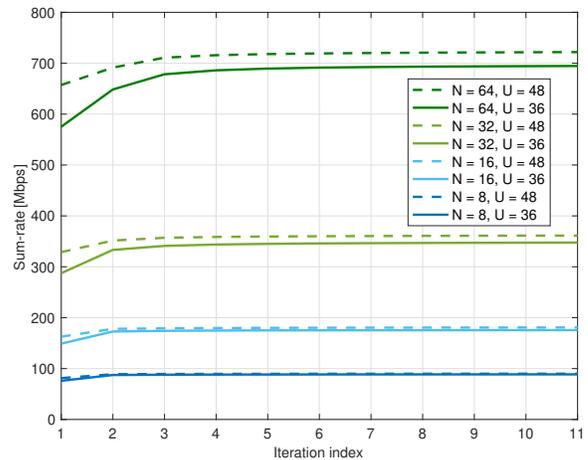}%
 \label{fig: ch4converge_inner}}
\hfil
\subfloat[]{\includegraphics[width=.95\linewidth]{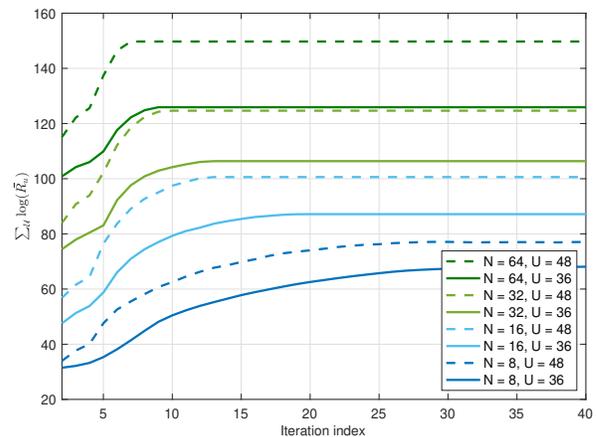}%
  \label{fig: ch4converge_outer}}
\caption{Convergence of sum-log average rate in Algorithm \ref{alg:ch4}. (a) Convergence of the inner loop, $\sum_u \alpha_u R_u$, in one proportional fairness index. (b) Convergence of the outer loop.} 
\label{fig: ch4converge}
\end{figure}

 Fig.~\ref{fig:ofdma_2scen} illustrates the performance of a cloud-enabled HAPS system for different user densities. First, the fairness across user services is illustrated in Fig.~\ref{fig:jain_2scen} as a function of the number of frequency tones/subcarriers and user density. Although the fairness measure of low user density is higher than in the high-density case, the difference between the two shrinks as we increase the number of resources. Second, Fig.~\ref{fig:users_2scen} demonstrates the percentage of served users in a high and low-density deployment of users as a function of $N$. Similar to the fairness index, in a low-density scenario, Algorithm \ref{alg:ch4} offers to serve a high percentage of users. This percentage drops, as expected, as we increase the user density since the number of available physical resources remains fixed.

Finally, Fig.~\ref{fig: ch4converge} illustrates the convergence of Algorithm \ref{alg:ch4}. Particularly, Fig.~\ref{fig: ch4converge_inner} presents the inner loop (i.e., the weighted sum-rate) convergence, and Fig.~\ref{fig: ch4converge_outer} shows the convergence of the outer loop (i.e., the sum-log of $\bar{R}$). 
Fig.~\ref{fig: ch4converge_inner} highlights the gain in the sum-rate obtained via optimizing the beamforming vectors, where the sum-rate improves by $15-20 \%$ compared to the initial value. On the other hand, Fig.~\ref{fig: ch4converge_outer} highlights the gain in the sum-log of the long-term average rate over $T = 60$ iterations. As Fig.~\ref{fig: ch4converge_outer} illustrates, increasing the number of tones $N$ (i.e., the number of available resources) achieves faster convergence to a  proportionally fair scheduling scheme. Overall, Fig.~\ref{fig: ch4converge_outer} shows that Algorithm \ref{alg:ch4} improves the proportional fairness in terms of the long-term average rate of users by approximately 50\%.

\section{Conclusion}\label{sec:conc}
While the development of previous generations of wireless communication networks focused on the need for higher data rates, the agenda for 6G and beyond networks focuses on digital inclusion.
In this direction, the research communities in the wireless communication area have joined their efforts to investigate ways of democratizing connectivity as a service. The limitation of terrestrial networks in providing global and equitable coverage motivates the deployment of non-terrestrial networks. This work focuses on augmenting terrestrial communication networks with a cloud-enabled high-altitude platform station to provide higher coverage for rural areas while ensuring sufficient capacity for hyper-digitized areas. The prolific potential of high-altitude platforms for computation and communication services supports such a connectivity framework. To this end, this paper investigates two distinct optimization problems that offer throughput maximization and load-balancing resource management schemes in cloud-enabled HAPS-assisted wireless communication networks. This paper studies the problem of designing the downlink association strategy and the corresponding beamforming vectors for throughput maximization and fairness improvement subject to transmit-power and fronthaul capacity constraints. The paper proposes tackling such complex mixed variable non-convex optimization problems using FP and sparse-beamforming techniques to obtain iterative numerical optimization algorithms. The proposed sum-rate maximization problem exhibits high throughput gains from augmenting terrestrial networks with hot-air balloons in cloud-enabled HAPS networks. On the other hand, the fairness-improvement optimization framework for load-balancing in cloud-enabled HAPS systems provides a proportionally-fair prioritization in scheduling users to resources. The proposed iterative optimization algorithm indeed accommodates more users and demonstrates remarkable capabilities in providing an inclusive digital framework,  which is perfectly aligned with future 6G networks theme at connecting the unconnected and ultra-connecting the connected. In fact, both the proposed space-air-ground system and the adopted sophisticated optimization design techniques in this paper offer one step forward toward enabling large-scale connectivity by means of intelligently integrating computation and communication from the sky, a topic that falls at the forefront of future wireless communication systems.

\bibliographystyle{IEEEtran}
\bibliography{References}

\end{document}